\documentclass[fleqn,usenatbib]{mnras}

\usepackage{newtxtext,newtxmath}

\usepackage[T1]{fontenc}

\DeclareRobustCommand{\VAN}[3]{#2}
\let\VANthebibliography\thebibliography
\def\thebibliography{\DeclareRobustCommand{\VAN}[3]{##3}\VANthebibliography}

\usepackage{graphicx}	
\usepackage{amsmath}	
\usepackage{xcolor}
\usepackage{multirow}

\renewcommand{\deg}{\hbox{$^\circ$}}
\renewcommand{\arcmin}{\hbox{$^\prime$}}
\renewcommand{\arcsec}{\hbox{$^{\prime\prime}$}}
\newcommand\itemMC{\item[\textbf{Class MC:}]}
\newcommand\itemnMC{\item[\textbf{Class S:}]}

\title[LOFAR Multi-modal classifier]{Identification of multi-component LOFAR sources with multi-modal deep learning}

\author[Alegre, Best, Sabater et al.]{Lara Alegre,$^{1}$\thanks{E-mail: alegre@roe.ac.uk}
Philip Best,$^{1}$
Jose Sabater,$^{1,2}$
Huub Rottgering,$^{3}$
Martin Hardcastle,$^{4}$
Wendy Williams$^{5}$
\\
\\
$^{1}$SUPA, Institute for Astronomy, University of Edinburgh, Royal Observatory, Blackford Hill, Edinburgh, EH9 3HJ, UK\\
$^{2}$UK Astronomy Technology Centre, Royal Observatory, Blackford Hill, Edinburgh, EH9 3HJ, UK\\
$^{3}$Leiden Observatory, Leiden University, PO Box 9513, NL-2300 RA Leiden, The Netherlands\\
$^{4}$Centre for Astrophysics Research, Department of Physics, Astronomy and Mathematics, University of Hertfordshire, College Lane, Hatfield AL10 9AB, UK\\
$^{5}$SKA Observatory, Jodrell Bank, Lower Withington, Macclesfield, SK11 9FT, UK\\
}

\date{Accepted XXX. Received YYY; in original form ZZZ}

\pubyear{2024}

\begin{document}
\label{firstpage}
\pagerange{\pageref{firstpage}--\pageref{lastpage}}
\maketitle

\begin{abstract}

Modern high-sensitivity radio telescopes are discovering an increased number of resolved sources with intricate radio structures and fainter radio emissions. These sources often present a challenge because source detectors might identify them as separate radio sources rather than components belonging to the same physically connected radio source. Currently, there are no reliable automatic methods to determine which radio components are single radio sources or part of multi-component sources. We propose a deep learning classifier to identify those sources that are part of a multi-component system and require component association on data from the LOFAR Two-Metre Sky Survey (LoTSS). We combine different types of input data using multi-modal deep learning to extract spatial and local information about the radio source components: a convolutional neural network component that processes radio images is combined with a neural network component that uses parameters measured from the radio sources and their nearest neighbours. 
Our model retrieves 94 per cent of the sources with multiple components on a balanced test set with 2,683 sources and achieves almost 97 per cent accuracy in the real imbalanced data (323,103 sources). The approach holds potential for integration into pipelines for automatic radio component association and cross-identification. Our work demonstrates how deep learning can be used to integrate different types of data and create an effective solution for managing modern radio surveys.

\end{abstract}


\section{Introduction}
\label{sec:intro}

The role of active galactic nuclei (AGN) in galaxy evolution is widely recognised today \cite[see reviews by][and references therein]{fabian2012observational, kormendy2013coevolution, heckman2014coevolution}, with AGN feedback being the main candidate responsible for suppressing star formation and leading to massive galaxies becoming ``red and dead''. Radio-loud AGNs, or radio AGNs for short, which have relativistic jets extending tens or hundreds of kiloparsecs from the galaxy, are thought to be the primary force behind this AGN feedback \citep[see][for a review]{Hardcastle2020}. However, certain aspects of AGN-galaxy co-evolution, such as the mechanisms by which AGNs are triggered, are not completely understood, and larger samples of AGNs are needed to permit detailed statistical studies \citep[e.g.][]{best2006, best2007, Sabater2019}. 

Significant advances have been made with data from extensive radio continuum surveys, such as the Faint Images of the Radio Sky at Twenty centimeters survey \citep[FIRST;][]{Becker1995}, the National Radio Astronomy Observatory (NRAO) Very Large Array (VLA) Sky Survey \citep[NVSS;][]{condon1998nrao} and the LOw Frequency ARray \citep[LOFAR;][]{vanHaarlen2013lofar} Two-meter Sky Survey \citep[LoTSS;][]{Shimwell2017,Shimwell2019lofar,Shimwell2022}. These surveys cover wider and deeper areas of the sky, which result in an increase in detected sources from tens of thousands in early radio surveys to about 5 million currently.
These surveys already provide large enough samples for some statistical studies, but with upcoming telescopes like the Square Kilometre Array \citep[SKA;][]{dewdney2009square}, it is anticipated that we will get a fully detailed picture from the radio viewpoint of galaxy evolution, AGN triggering, and the influence of AGNs on galaxies across cosmic time. However, to perform these studies, it is crucial to obtain accurate measurements of radio fluxes and source sizes in order to characterise the radio AGN properties of the hosting galaxy. Furthermore, it is necessary to have precise identification of the radio source host galaxy to obtain optical properties, as well as redshifts to enable measurements to be converted into physical properties. 

In LoTSS, radio source properties including source sizes and flux densities are measured using the Python Blob Detector and Source Finder \citep[\texttt{PyBDSF};][]{Mohan2015pybdsf}, which extracts confined regions of high radio brightness from the images, designated as \texttt{PyBDSF} sources, which can be fitted by one or various Gaussians.
In order to get the correct optical counterparts, in LoTSS DR1, a proportion of the \texttt{PyBDSF} sources were visually inspected while the majority of them were cross-matched automatically using the statistical Likelihood Ratio (LR) technique \citep[see][]{williams2019lofar}.  
When sources were visually inspected, they fell mainly into three categories. 
The first category was extended single-component radio sources. These are sources that have been successfully identified as physical sources by \texttt{PyBDSF}. However, due to their extended radio emission, automatic cross-matching methods become less reliable, requiring visual inspection and cross-identification. Machine-learning methods have been developed that show promising potential for providing accurate cross-match IDs for these type of sources. For example, \citet[]{Alger2018} implemented a method that involves creating a bounding box centred on a radio component and deriving a score for potential candidate IDs within a search radius, demonstrating significant efficacy in cross-matching sources of this nature.
The second category of sources comprises blended sources, where \texttt{PyBDSF} encompasses multiple sources into a single detection, necessitating deblending before cross-matching. As demonstrated, for example by \citet[][]{williams2019lofar}, the implementation of automated algorithms for source deblending can be accomplished with relative ease.
Thirdly, there are radio sources composed of multiple components (MC). In these cases, \texttt{PyBDSF} separated a physical radio source into different source components, and it is therefore necessary this category to associate the components before cross-matching.
 
MC sources are typically sources with extended radio emission and/or distinct radio blobs. When applying source detection algorithms \citep[e.g.][]{Mohan2015pybdsf, Hale2019} to high-resolution images, algorithms search for pixel areas exceeding a pre-determined threshold level (often set at a signal-to-noise ratio of 5). Sometimes certain parts of a source may fall below the threshold level, and therefore the software may identify different source regions above the threshold as separated sources. Extreme cases are FRIIs \cite[see][for FRI vs. FRII classification]{fanaroffriley1974}, which possess highly luminous steep-spectrum lobes but faint flat-spectrum jets between the lobes, which commonly fall below the signal-to-noise level. Sometimes, even if detections are above the threshold level, it is possible for certain components to be separated as the software tries to remove irrelevant sources to avoid incorrectly producing blends. 

The cross-identification of MC sources presents a significant challenge, since it involves the accurate definition of the radio source (which requires radio source component association) and the cross-matching of the (potentially very extended) radio source to its optical counterpart.
Some algorithms have recently been developed to group components of MC sources in radio images \citep[e.g.][]{wu2019Claran, Mostert2022}, and others successfully identify the host galaxy in source components that have already been grouped beforehand \citep{barkus2022application}.
However, without a specific methodology, it is impossible to determine whether a source requires component association. When applying these algorithms without previous knowledge then if, for example, the source needs component association, a bounding box, which may encompass only one of the source components, may give the correct ID \citep[e.g.][]{Alger2018}, but the radio source properties will be incorrect. Consequently, the initial step of cross-matching MC sources involves ensuring the appropriate identification of a source as a MC source in order to determine whether the radio components have been accurately associated or not.

Due to their complexity, MC sources hold significant interest for both individual galaxy studies \citep[e.g.][]{Hardcastle2019X} and statistical studies \citep[e.g.][Alegre et~al., in prep.]{Sabater2019, Hardcastle2019DR1}. Hence, it is crucial to identify these sources to precisely measure their radio properties.  
Given the lack of automatic methods available for identifying MC sources, our primary focus in this paper is to identify them.
To address this we use Machine Learning (ML) and the LoTSS data. We employ Multi-Modal ML \citep[MML; e.g.][]{ngiam2011multimodal}, a ML type of model that integrates different data inputs. In MML models, each data instance can contain various types of information, such as images, structured data, and others such as text, audio, video, and even metadata \citep[see, e.g.][]{baltruvsaitis2018multimodal}.

MML has been successfully applied to a wide range of AI problems, with particular developments in deep learning and computer vision \citep[see][and references therein]{summaira2021recent}. However, MML methods have only recently been developed to be used in astronomy applications. For example, \cite{hong2023photoredshift} used a MML model to estimate photometric redshifts of galaxies in the Sloan Digital Sky Survey \citep[SDSS,][]{York2000} with significant improvement in the estimations. Natural language processing combined with radio images from the ``Radio Galaxy Zoo: EMU'' were recently used to classify galaxies based on description tags \citep{Bowles2023mm}.
In the context of weak gravitational lensing, \cite{vago2022deepgravilens} combined images and time-series data to detect lensing effects in four different simulated survey datasets, showing that the method surpasses the traditional method using only images, which will be important to detect lenses in upcoming surveys, such as the Large Survey of Space and Time \citep[LSST;][]{Ivezic2019lsst}. \cite{cuoco2021multimodal} combined information from different parts of the electromagnetic spectrum to characterise gravitational wave events. They further reviewed the computational aspects of MML astrophysics and the importance of developing methods that combine multimessenger astronomy \citep[][]{cuoco2022computational} . The complexity and amount of data that new gravitational wave detectors and new telescopes will generate by detecting thousands of transients per night, creates urgency for developing methods that are able to efficiently analyse and combine the information coming from multiple sources. 

In this work, we train a MML classifier built on a Convolutional Neural Network (CNN) and an Artificial Neural Network (ANN) in order to identify MC sources. The MML model combines two different types of information into a unified architecture; it takes as inputs radio and optical properties as well as radio images. Although somewhat similar architectures have been used in radio astronomy \citep[e.g. by connecting 2 CNNs,][]{Viera2021, samudre2022data} these approaches do not combine data coming from different sources, or different data types. 
By incorporating multiple sources of information, in this paper we demonstrate improved performance in identifying MC sources in LoTSS and show the advantages of using MML to analyse future radio surveys. Furthermore, we employ active learning \citep[e.g.][]{Walmsley2020active}, by using the results from \cite{alegre2022} to remove sources from the dataset that are less informative for the learning process. By selecting the most informative sources for the model, it is possible to optimise its performance while also reducing the number of examples needed for training.

The paper is organised as follows. We describe the LoTSS data and the creation of the dataset in Section~\ref{sec:data}, where we define the data types, define the classes and discuss how balancing the dataset was achieved. We then perform a set of experiments in Section~\ref{sec:experiments}. We define a baseline model and explore the production of the images, the creation of the multi-modal model (where we test for different sets of features), data augmentation, as well as adjusting the training dataset. We further present the model optimisation and model performance. The model is applied to the real imbalanced data sample in Section~\ref{sec:fulldataset}. We conclude and discuss future directions in Section~\ref{sec:conclusions}.

\section{Data}
\label{sec:data}

This work is focused on data from the LoTSS survey carried on with the LOFAR telescope. LoTSS is a survey of the entire northern sky which reaches depths about 10 times greater than the FIRST survey (for sources of typical steep spectral index), while achieving sensitivity to extended structures, better than the NVSS survey. This unique combination allows for the detection of sources with extended faint emission. LoTSS has a frequency coverage from 120 to 168 MHz, and achieves a typical rms noise level of 70~$\umu$Jy/beam over its first data release (DR1) region, with an estimated point source completeness of 90 per cent at a flux density of 0.45\,mJy. The low frequencies of LOFAR combined with high sensitivity on short baselines gives it a high efficiency at detecting extended radio emission. LoTSS DR1 has an angular resolution of 6\arcsec\ and an astrometric precision of 0.2\arcsec, making it very suitable for host-galaxy identification. In this section, we provide an overview of the LoTSS DR1 data and the dataset that is extracted from it to perform the experiments. More details about the data used to create the dataset can be found in \cite{alegre2022}.

\subsection{LoTSS} 

LoTSS detected 325,694 \texttt{PyBDSF} sources in its first data release, containing just the first 2 per cent of the survey \citep[LoTSS DR1;][]{Shimwell2019lofar}~\footnote{\hyperlink{https://lofar-surveys.org}{https://lofar-surveys.org}}. The public release provided radio catalogues that were derived from the 58 mosaic images of DR1, which cover 424\,deg$^2$ over the Hobby-Eberly Telescope Dark Energy Experiment \citep[HETDEX;][]{hill2008hobby} Spring Field (right ascension 10h45m00s -- 15h30m00s and declination 45\deg00\arcmin00\arcsec-- 57\deg00\arcmin00\arcsec). The area benefits from extensive multi-wavelength coverage. The released LoTSS data products include value-added catalogues which present the identification of LOFAR-matched radio sources to optical counterparts using Pan-STARRS \citep{Chambers2016} and the Wide Infrared Survey Explorer \citep[WISE,][]{Cutri2013AllWISE} surveys, achieved using a combination of statistical techniques and visual inspection via a private LOFAR Galaxy Zoo (LGZ) classification project hosted on the Zooniverse platform\footnote{\hyperlink{https://www.zooniverse.org}{https://www.zooniverse.org}} \citep[described in paper III of LoTSS DR1,][]{williams2019lofar}. The catalogues also provide some initial characterisation of the sources, including photometric redshift estimates and rest-frame magnitudes \citep[described in paper IV of LoTSS-DR1;][]{Duncan2019lofar}.

In LoTSS DR1, sources bigger than 15 arcseconds (that were not automatically cross-matched with a large SDSS optical source) were all sent to visual inspection without any triage, since large sources are usually resolved and potentially complex. These correspond to 19,216 sources, or 5.95 per cent of LoTSS DR1. From these, the outcome of the visual analysis  demonstrated that 10,001 (52.05 per cent) needed to have been inspected \citep{alegre2022}, with 4,671 (24.31 per cent) being MC sources and the rest single-component sources. Considering only the large and bright sources (total flux $>$ 10 mJy, 6,748 sources, 2.09 per cent of LoTSS DR1), i.e. the ones used to perform component association by \citet{Mostert2022}, only 2,226 (32.99 per cent) of those in fact needed component association, decreasing the performance of source association. Even though the majority of the components of MC sources are indeed large and bright, the remaining components fall into different parts of the \citet{williams2019lofar} decision tree, with only 57 MC sources (0.63 per cent) being sent directly to visual inspection, 201 (2.22 per cent) being automatically cross-matched with a large optical galaxy but inspected afterwards, 1,046 (11.56 per cent) being accepted automatically to cross-match by LR (most likely the cores of FRII or double-lobed sources), and finally 3,071 (33.95 per cent) going through a pre-filtering process before further visual analysis. 

A second LoTSS data release with a total number of 4,396,228 \texttt{PyBDSF} sources in 841 mosaics covering 5634~deg$^2$ has been published \citep[LoTSS DR2; ][]{Shimwell2022}; some aspects will be discussed in Section~\ref{sec:4mjy}. LoTSS DR2 corresponds to 27 per cent of the northern sky and it spans two regions: one with 4178~deg$^2$ around right ascension 12h45m and declination 44$\deg$30\arcmin  and the other with 1457~deg$^2$ around right ascension 1h00m and declination 28$\deg$00\arcmin. LoTSS DR2 has a central frequency of 144~MHz with 83~$\umu$Jy/beam rms sensitivity and an estimated  point-source completeness of 90 per cent at a peak brightness of 0.8~mJy/beam. \cite{Hardcastle2023dr2} present the methods used to cross-match LoTSS DR2 radio sources with their corresponding optical counterparts. In their work, the public Zooniverse `Radio Galaxy Zoo: LOFAR' was established for the purpose of associating and cross-matching a fraction of the sources in the dataset.

\subsection{Dataset classes}
\label{sec:classes}

In this work, we use supervised machine learning for classification, which involves training models using labelled data with the aim of classifying unseen examples afterwards. The labelled data provided for training determine the quality of the model and its ability to generalise (i.e. to be able to classify other examples correctly). Therefore, it is important to have a well-defined and well-annotated dataset. We created the dataset using 323,103 \texttt{PyBDSF} sources, which resulted from removing the artefacts from the original 325,694 \texttt{PyBDSF} sources obtained over the LoTSS DR1 area. This was done by comparing the original \texttt{PyBDSF} radio catalogue with the outputs obtained from a combination of visual inspection and statistical cross-matching described in detail in \cite{williams2019lofar}. In cases where source components had been merged, this resulted in a single entry in the final catalogue; deblended sources, on the other hand, show multiple entries. Single-component sources remain the same in both catalogues. This enables the categorisation of sources into two distinct classes: class MC corresponds to multi-component (MC) sources, whereas class S is a mix of non-MC sources.~\footnote{The supplementary online material includes a list of the 323,103 \texttt{PyBDSF} sources, with sources in class MC assigned a value of 1 in the \textit{multi$\_$component} column and sources in class S assigned a value of 0.}  They are defined as follows:

\begin{enumerate} \addtolength{\itemindent}{0.8cm}
    \itemMC \noindent \texttt{PyBDSF} sources that were associated with other \texttt{PyBDSF} sources in LGZ, meaning that these make up a MC source. These correspond to sources for which the \texttt{PyBDSF} algorithm has detected the radio emission separately, or has split the radio emission, giving rise to two or more different radio components. To construct a genuine physical source it is, therefore, necessary to associate the different source components.
    \itemnMC \noindent \texttt{PyBDSF} sources for which the source emission is all encompassed within a single \texttt{PyBDSF} source, and therefore do not require component association. While these primarily consist of correctly identified single-component sources, this class also includes the blended sources that \texttt{PyBDSF} incorrectly identified as being a single source and that needed to be split into two or more sources.
\end{enumerate}

Artefacts correspond to \texttt{PyBDSF} sources not present in the final LoTSS DR1 value-added catalogue and have been excluded from this analysis.

\subsection{Balancing the dataset}

In radio surveys the number of objects in the two different classes is highly imbalanced, with relatively low numbers of class MC sources. Balancing the dataset (having a similar number of examples in each class) is a common ML technique used to avoid overfitting the model to the majority class during the training process. A balanced dataset was achieved using an undersampling method, which consists of using only a subsample of all the available data. This has been shown to be effective by \citet{alegre2022}. 
Furthermore, the augmentation step (adding more examples through rotations and reflections; see Section~\ref{sec:augmentation}) will act as compensation for the undersampling, whereby the class MC sources will be effectively augmented while different groups of class S will be added without suffering augmentation transformations. Consequently, a greater number of examples will be used to train this algorithm since, typically, deep learning algorithms require more training data than machine learning ones.
It is worth noting that the model is trained using a balanced dataset but is it then used to make predictions on data that has an unequal distribution of classes. Particular attention must be paid to this when applying the classifier to real distributions (see Section~\ref{sec:fulldataset}).

The balanced dataset (before augmentation) has 9,046 sources in each class. Class S corresponds to 8,189 random single sources and 857 blended \texttt{PyBDSF} sources, which were included in the dataset class because even if they are rare, they will be part of the real datasets, and thus, they allow the classifier to train using a wider variety of single sources. 
Class MC consists of 9,046 multi-component \texttt{PyBDSFs}, which is reduced from the 9,072 sources that required component association, as 26 sources were both deblended and grouped with another \texttt{PyBDSF} source, and therefore were excluded because they belong to distinct classes.
The total number of sources in the balanced dataset before augmentation is 18,092, with the training set corresponding to 12,664 sources (70 per cent) and the validation and test sets corresponding to 2,714 sources each (15 per cent each).

\subsection{Dataset images}

The images used to create the dataset are cutouts around the \texttt{PyBDSF} sources centred on their right ascension and declination positions. These were cut from the 58 LoTSS DR1 mosaics~\footnote{\hyperlink{https://lofar-surveys.org/dr1$\_$release.html}{lofar-surveys.org/dr1$\_$release}}. The DR1 mosaics have 1.5 arcsec/pixel, and the final images used are 128$\times$128 pixels PNGs, corresponding to 192$\times$192 arcseconds. The images were, however, extracted first as 256$\times$256 pixel FITS files, which were then used for augmentation (including rotation -- hence the larger size to avoid empty regions in the corners after rotation; see Section~\ref{sec:augmentation} for details), application of sigma cuts, and combining the different channels, before being cut into the final 128$\times$128 pixel images.
Initially, a default sigma clipping on a linear range between 1$\sigma$ and 30$\sigma$ was applied to the images. Different authors \citep[e.g.][]{Aniyan2017, alhassan2018first, tang2019transfer, Mostert2022} have shown that the performance of a CNN model depends on the background noise in the input images. Thus, later, we will also investigate different sigma cuts that enhance extended emission while simultaneously removing noise. In all cases, the PNG images were normalised after applying the sigma cuts, in which the values were scaled to a range of 0-1. This makes it easier to create composite images and also reduces computational costs when using a 3-channel CNN. More details about preprocessing the images can be found in Section~\ref{sec:sigmacuts}.

Although the entire source may occasionally (but rarely) extend outside the frame, the choice of an output image size of 128$\times$128 pixels is a reasonable compromise. There are only 199 final associated LGZ sources in the sample for which the angular size is larger than the picture frame. This represents only 6 per cent of the total final associated MC sources (3,596 sources), and even for these, each of the source components is significantly smaller than the image size chosen. In all these cases, there is still a substantial quantity of information within the frame. The objective is only to determine whether or not the source is a MC, not to identify all of the source components. Therefore, a source being larger than the image does not represent an issue, as it would potentially be if we were conducting tasks such as source association, morphology classification, or host galaxy cross-matching using ML. Furthermore, the classifier does not necessarily need the image of the entire source to determine that it requires association, with the extended emission being a better indicator. 

\section{Constructing the multi-modal model}
\label{sec:experiments}

In this section, we conduct experiments that will ultimately lead to the adoption of a final model. A Convolutional Neural Network is investigated in Section~\ref{sec:CNN} where a baseline architecture (Section~\ref{sec:baseline}) is established to allow for the evaluation of changes made to various aspects of the model, with further experiments examining image production (Section~\ref{sec:sigmacuts}). The extension to a multi-modal architecture is explained in Section~\ref{sec:multimodel}, with adjustments made to the training set described in Section~\ref{sec:nsisg} and augmentation in Section~\ref{sec:augmentation}. Modifications to model hyperparameters are described in the optimisation stage (Section~\ref{sec:optimisation}); and the final model performance is presented in Section~\ref{sec:performance}.
The evaluation criteria used for assessing the performance of the model are explained in Appendix~\ref{sec:ML_metrics}.

\subsection{The Convolutional Neural Network}
\label{sec:CNN}

\subsubsection{Establishing a baseline model}
\label{sec:baseline}

In order to establish a baseline model, we assessed different convolutional neural networks (CNN) that had been used for radio morphology classification. These correspond to models developed mainly for classifying sources into FRI, FRII, bent-tailed, and compact sources \citep[][]{alhassan2018first, Aniyan2017, becker2021cnn}, but also to differentiate between compact and extended sources \citep{lukic2018radio}.
These models\textbf{} are expected to provide a good starting point for the experiments. All of the models we are testing here were originally developed to work with VLA FIRST data. The higher radio frequency of FIRST (1.4 GHz) results in distinct regions of the observed galaxies being more prominent in FIRST images compared to the ones in the LOFAR surveys. Particularly, FIRST emphasises galaxy cores and hotspots, whereas LoTSS (144 MHz) provides a broader picture of the source, highlighting much more extended emission. However, this also suggests that the architectures under consideration may be suitable for the current task, as they have demonstrated efficacy in identifying sources that appear as separated radio emission (e.g. FRII) in particular for the FIRST data and therefore may be useful to identify MC sources.

Here, we provide an overview of the models we have chosen to investigate and any modifications we have made to the architectures and hyperparameters. The corresponding publications \citep[][]{alhassan2018first, Aniyan2017, becker2021cnn, lukic2018radio} provide detailed descriptions and illustrations of the structure of each model.
All tested architectures have a set of convolutional layers, typically each made up of a convolutional stage (where feature extraction is performed, outputting feature maps), a detection stage (based on a non-linear activation function, commonly a Rectified Linear Unit, or ReLU), and a pooling stage (which subsamples the feature maps, reducing their spatial size, in this case by using maxpooling, where the maximum values are retained). Following this, the architectures have a final 1 to 3 dense layers with dropout (a regularisation technique that corresponds to removing random neurons), followed by a softmax layer (which transforms the outputs into probabilities). A kernel is applied to the input image and the feature maps during the convolutional and pooling operations. This operates as a sliding window, computing dot products between the kernel values and the values of the pixels of the images and feature maps in the convolutional stages and retaining specific values in the pooling stage (e.g. the maximum value). Different strides can be applied, where the stride corresponds to how many steps the kernel shifts in the horizontal and vertical directions after each computation.
Smaller kernels and strides mean tighter scanning, possibly enabling more details to be extracted from the images. These filters are learnable matrices that specialise in detecting different features, with a higher number of filters having the potential to identify increasingly complex and intricate patterns.

Both \cite{alhassan2018first} and \cite{lukic2018radio} have very similar architectures, with only 3 convolutional layers and 1 or 2 final dense layers, respectively, with 50 per cent dropout before the softmax layer. \cite{lukic2018radio} uses small sets of filters (16, 32, 64), while \cite{alhassan2018first} uses (32, 64, 94). \cite{alhassan2018first} uses kernels that are typically smaller and have smaller strides, while \citet{lukic2018radio} uses typically higher values for the kernel sizes and strides, in particular in the first layers.  Furthermore, \citet{lukic2018radio} uses two final dense layers of 1024 neurons each, while \cite{alhassan2018first} uses a single layer with only 194 neurons. Both models require a high number of epochs to converge. The \cite{alhassan2018first}  classifier was trained for 400 epochs and \cite{lukic2018radio} for 100 epochs. 

On the other hand, \cite{Aniyan2017} present a deeper (and wider) network, resulting in a much heavier model than the previous two architectures. With 5 convolutional layers (not all with a pooling stage), set for a large number of filters in each layer (96, 256, 384, 384, 256), three final wide dense layers with 4096 neurons each, and 50 per cent dropout, this model has a layer normalisation after the ReLU on each convolutional layer. The kernel sizes are larger in the first layers, with a stride of 1. For all of these reasons, this is an expensive model to run. The authors, when training it, ran it only for 30 epochs. In order to decrease memory problems, we had to make major changes to this architecture in particular. The filter sizes were reduced to (16, 32, 64, 64, 32) and the 3-dense layers to 1024 neurons each. Furthermore, the normalisation layer had to be removed because it was making the model unstable and leading to overfitting.

The \cite{becker2021cnn} model has a much deeper architecture but a lighter one as well. This model has 11 convolutional layers but only uses pooling every 3 (or 2) layers. It has a small number of filters in each set of three consecutive layers (32, 64, 128) and 256 on the two final layers, with a 25 per cent dropout after the maxpooling layer. Kernels have a general size of 3 and a stride of 1. It finishes with only one dense layer of 500 neurons and 50 per cent dropout, followed by a softmax. Most importantly, this original architecture required only 16 epochs to be trained.

We use a single-channel CNN to explore the different architectures and establish the baseline model. The model inputs radio images with a size of 128$\times$128 pixels, which went through a linear cut ranging from 1$\sigma$ to 30$\sigma$. No data augmentation was used at this stage.
The architectures were implemented with very few changes and assumptions, with only the \cite{Aniyan2017} architecture undergoing significant modifications, as explained above. Minor adjustments had to be made, particularly in cases where not enough details were provided. It was assumed that the stride corresponded to the size of the kernel, and padding was chosen to preserve the size of the feature maps (padding is used to add extra pixels with zero values to the border of the input image and feature maps before the convolutions). In cases where the specific location for the dropout layer was not explicitly indicated, dropout was performed after the dense layer, as per the original paper from \cite{Hinton2012dropout}.

The batch size, learning rate, optimiser algorithm, batch normalisation, and number of epochs were initially tested to find a suitable set of hyperparameters which allowed to compare the models without stability problems (huge variations across epochs) and massive overfitting issues. We tested each architecture individually using its original hyperparameters, but the classifiers overfitted or the performance was worse in general.
We found that the optimal hyperparameters for the four architectures were a learning rate of 0.0001, no batch normalisation (in architectures that applied it), a batch size of 32, and the use of the RMSprop optimiser \citep{tieleman2012lecture}. All the models were able to converge and show above 85 per cent accuracy after about 30 epochs of training. It was observed that the choice of these hyperparameters did not depend strongly on the architecture. We set these as the baseline hyperparameters.

\begin{figure}
\centering
\includegraphics[width=\columnwidth]{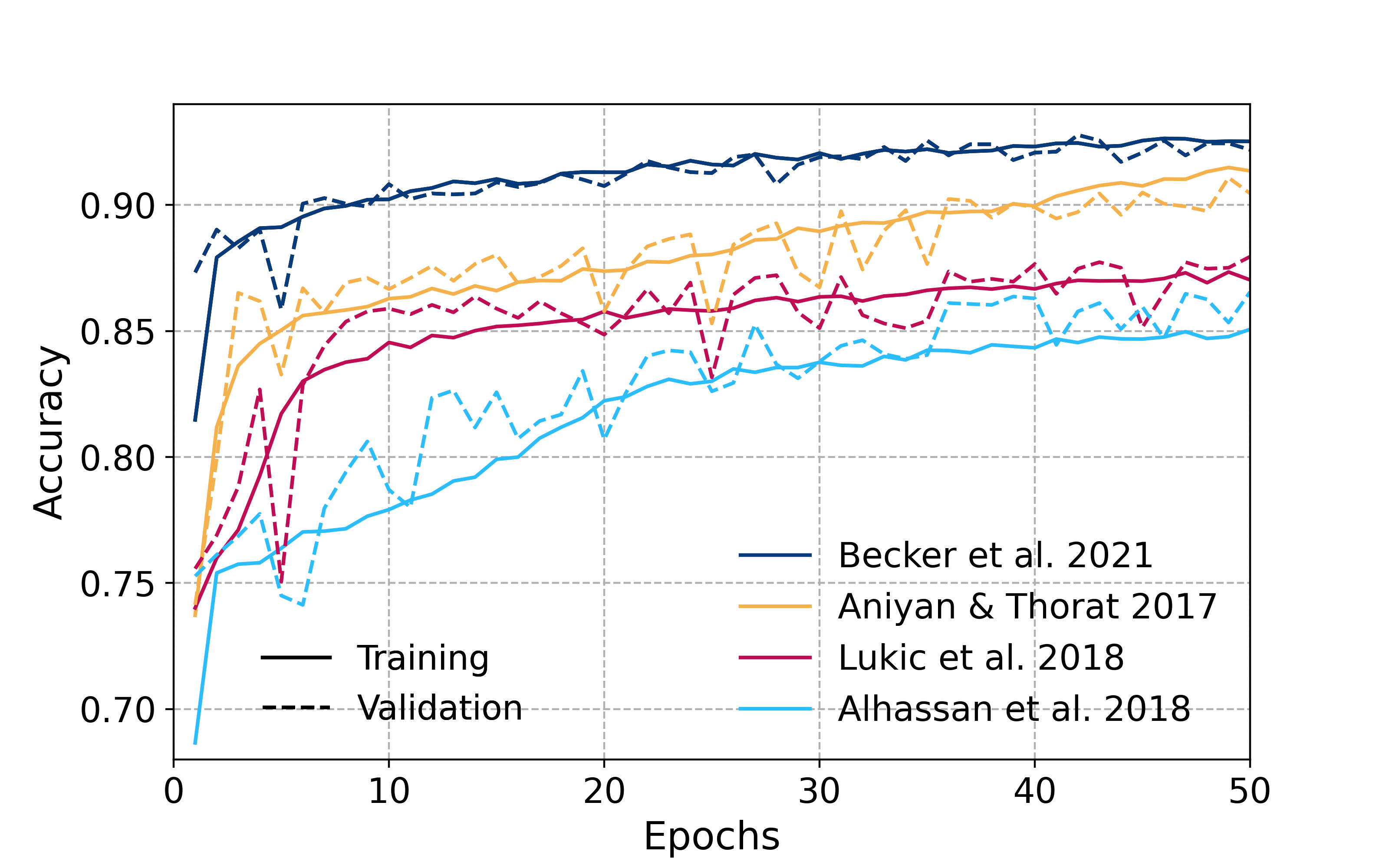}
\caption[Baseline experiments]{Performance of four different Convolutional Neural Networks (CNN) architectures that were used to establish the baseline model. All architectures were ran using the same set of hyperparameters (here showing training for 50 epochs), which were found to be the most suitable ones regardless of the CNN used (see text for a discussion).}
\label{fig:baseline}
\end{figure}

The performance of the different architectures on the training and validation sets is compared in Figure~\ref{fig:baseline} (see Appendix~\ref{sec:ML_metrics} for a definition of the performance metrics used). Even with small modifications made to the network and hyperparameters (for example, by introducing batch normalisation after each convolution layer or changing the learning rate to 0.001), the performance of the \cite{alhassan2018first} classifier is the weakest (reaching 85 per cent accuracy). The \cite{lukic2018radio} network performs about 2 per cent better and benefits from using a larger batch size and a smaller learning rate, which reduces the overfitting of the network when compared to using its original hyperparameters. Changes to the \cite{Aniyan2017} architecture resulted in good performance for the model, reaching accuracy values above 90 per cent, but the model shows some architectural issues, resulting in high training costs and also instabilities. For example, with the original learning rate of 0.01, the network was not even able to converge. Even though Figure~\ref{fig:baseline} suggests that it is possible that the model has the potential to improve its performance after more training, for the reasons mentioned (and also because there is a better alternative architecture), this model was excluded from further consideration. The model based on the \cite{becker2021cnn} architecture reaches accuracy values on both the training and validation sets above 92 per cent, and reducing the learning rate leads to even better results than the original one of 0.001. 

Overall, it is evident that, after establishing the baseline hyperparameters, the deeper architectures show superior performance for the identification of multi-component sources. The results of the model based on the \cite{becker2021cnn} architecture show the best performance, with similar values on both training and validation sets and high stability. This architecture performs well, converges rapidly, and trains smoothly. Therefore, it was selected as the baseline model. 

The hyperparameter values established for the baseline model will be the ones used throughout the experiments, unless stated otherwise, for example, when augmentation is introduced. 
The model which will be finally adopted is a refinement of this baseline model. The process of refinement and optimisation of both the hyperparameters and the architecture is described in Section~\ref{sec:optimisation}, which also contains a diagram illustrating the architecture of the final model.


\subsubsection{Optimising image production}
\label{sec:sigmacuts}

LoTSS original images show differences in noise levels depending on the sky regions being observed, and also show different contrast ranges with very bright sources or others with weak diffuse emission. We use sigma clipping for cleaning and removing noise from the images. This was done using Montage~\footnote{\hyperlink{http://montage.ipac.caltech.edu}{montage.ipac.caltech.edu}}, an astronomical image mosaic engine from NASA. The sigma-clipping procedure discards values (i.e. sets them to the minimum or maximum value) that are either above or below a defined standard deviation from the mean. 

\begin{figure*}
\begin{minipage}{\textwidth}
\normalsize{
\hspace{0cm} Original 
\hspace{1.2cm} 1$\sigma$-200$\sigma$ log (1) \hspace{1.2cm} 3$\sigma$ (2) 
\hspace{1.2cm} 1$\sigma$-30$\sigma$ log (3)
\hspace{1.2cm} (1, 2, 3)}
\centering
\includegraphics[scale = 0.19]{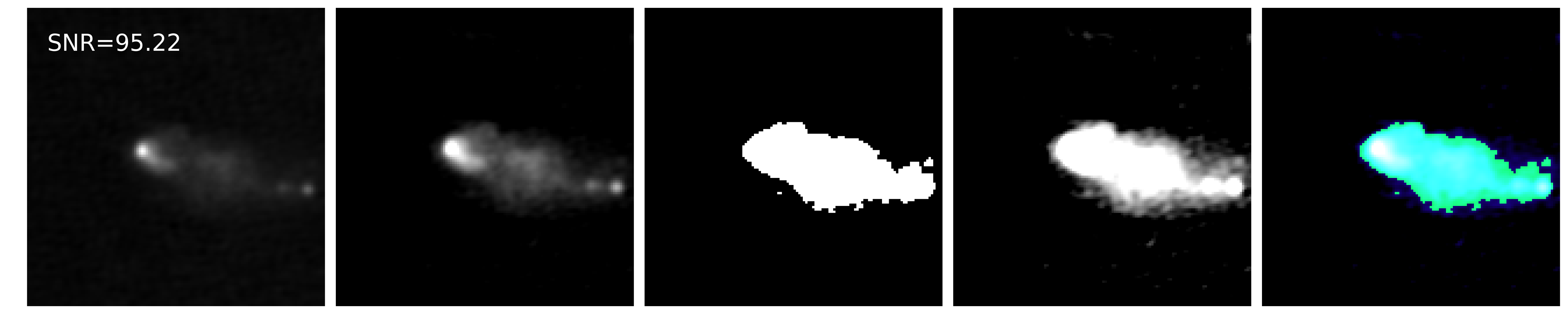}
\includegraphics[scale = 0.19]{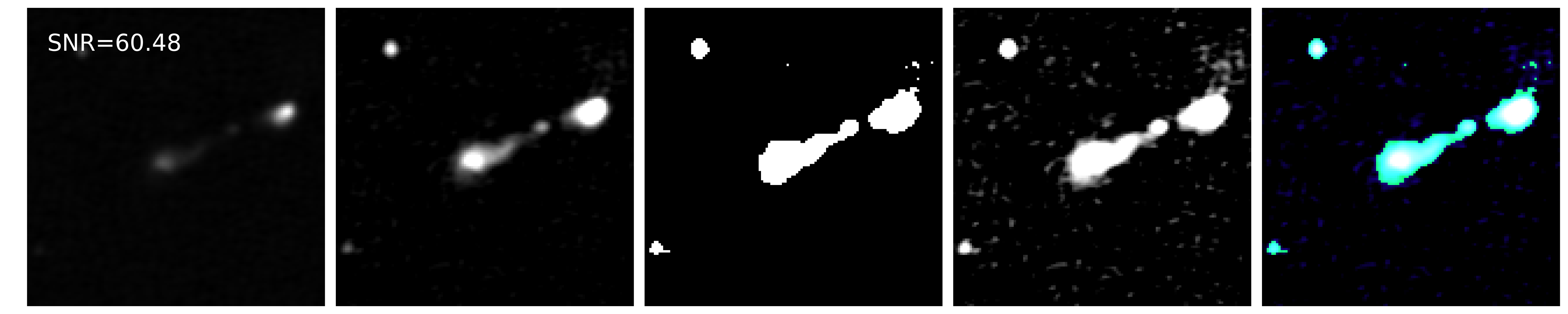}
\includegraphics[scale = 0.19]{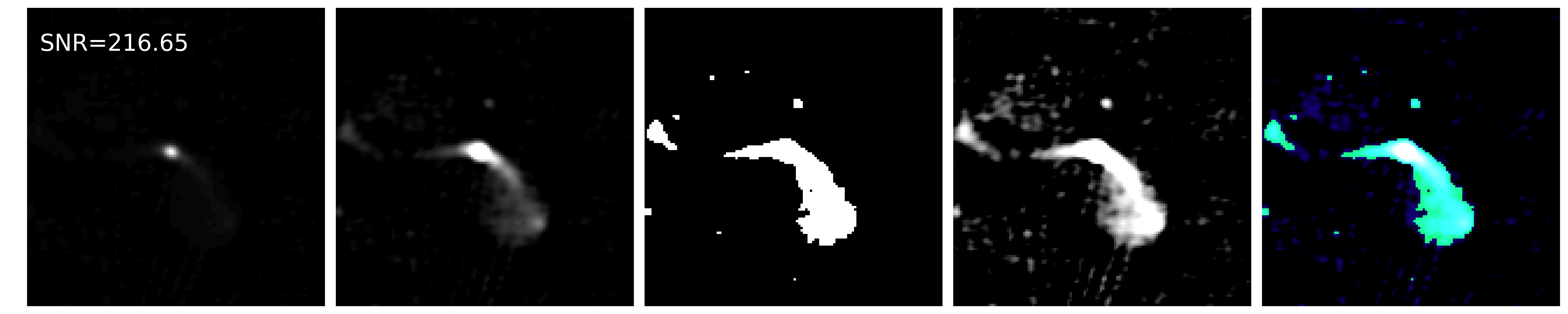}
\includegraphics[scale = 0.19]{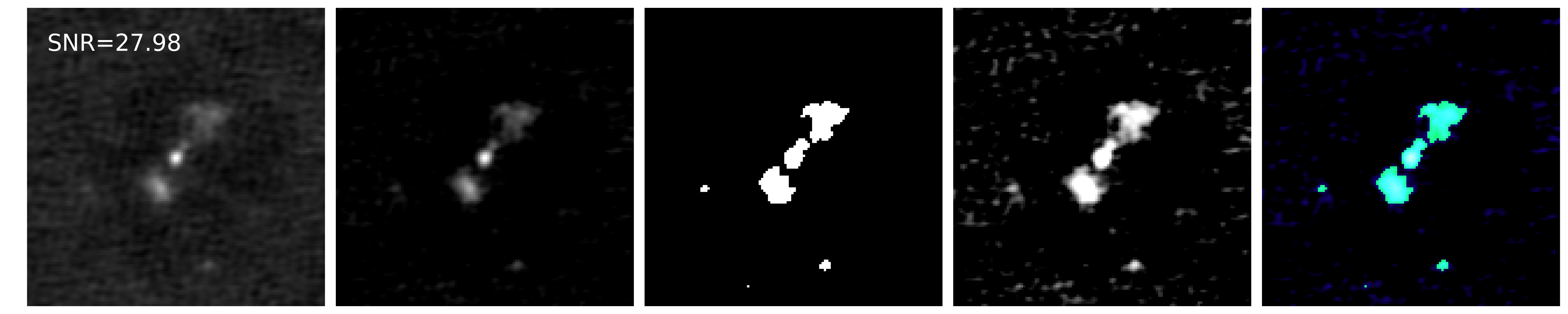}
\includegraphics[scale = 0.19]{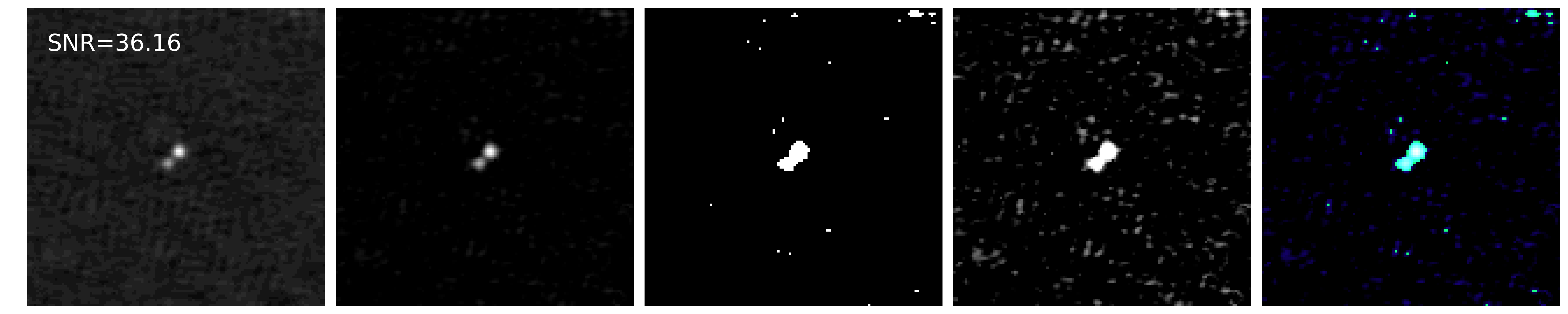}
\includegraphics[scale = 0.19]{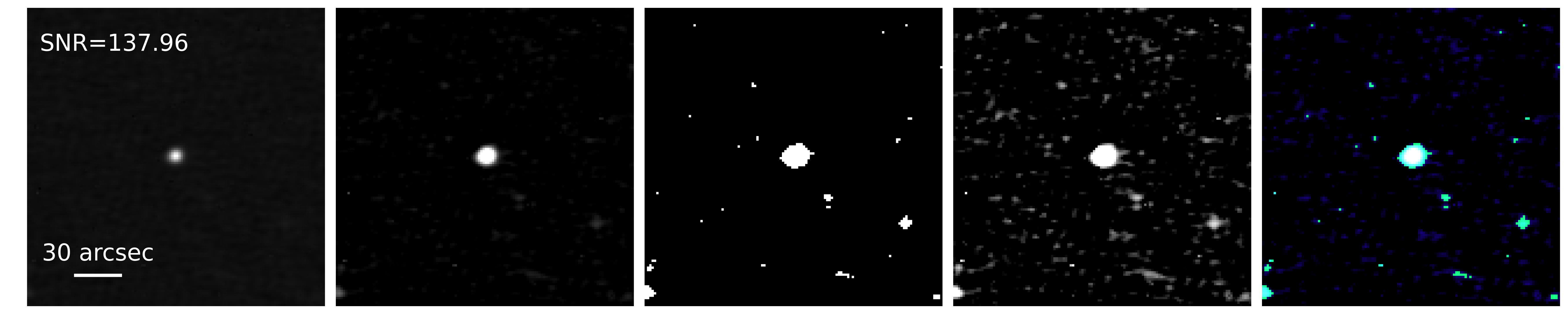}
\end{minipage}
\caption[Examples of sigma-clipping images]{Sigma-clipping image examples. The left column shows the original (not scaled) images directly extracted from the LoTSS DR1 mosaics, with indicated peak flux signal-to-noise ratio (SNR). The three middle columns correspond to individual sigma cuts, with black indicating the lower limit of the range and white indicating the upper limit. The right column is a composite image made up of the three individual ones, which is finally used in the 3-channel CNN. The first four rows correspond to multi-component sources. The fifth and sixth rows show a blended source and a single-component source, respectively. In the top row, the \texttt{PyBDSF} source corresponds to a lobe and the entire source is not within the frame; however, it is clear that enough source is present for the classifier to identify this as part of a MC source, justifying our choice of 128$\times$128 pixel image sizes even for the small fraction of sources that are larger than this.}
\label{fig:sigma_cut_example}
\end{figure*}

As baseline, we used image cuts of 1$\sigma$–30$\sigma$ on a linear scale. We also tested using sigma cuts of 1$\sigma$–30$\sigma$ and 1$\sigma$-200$\sigma$ in the logarithmic scale, and 3$\sigma$ and 5$\sigma$ cuts. In the wide range examples (1$\sigma$–200$\sigma$ and 1$\sigma$–30$\sigma$), the lower limit corresponds to 1$\sigma$, while the upper limit corresponds to 200$\sigma$ and 30$\sigma$, respectively, with a stretch applied on a logarithmic scale. The 3$\sigma$ (or 5$\sigma$) cut sets all values below 3$\sigma$ (or 5$\sigma$) to zero and sets values above that level to unity. Figure~\ref{fig:sigma_cut_example} compares these different sigma-clipping levels for some example sources. We can see that when using 1$\sigma$–200$\sigma$, the bright features and the diffuse emission of the source have been enhanced. Additionally, the extended emission has been smoothed out, and the background noise has been reduced, making it more consistent across images. The 3$\sigma$ cut displays the source silhouette in its entirety. The 1$\sigma$–30$\sigma$ emphasises the extended emission while maintaining a consistent level of noise across all images.

\begin{figure*}
\begin{minipage}{\textwidth}
\centering
\includegraphics[scale=0.47]{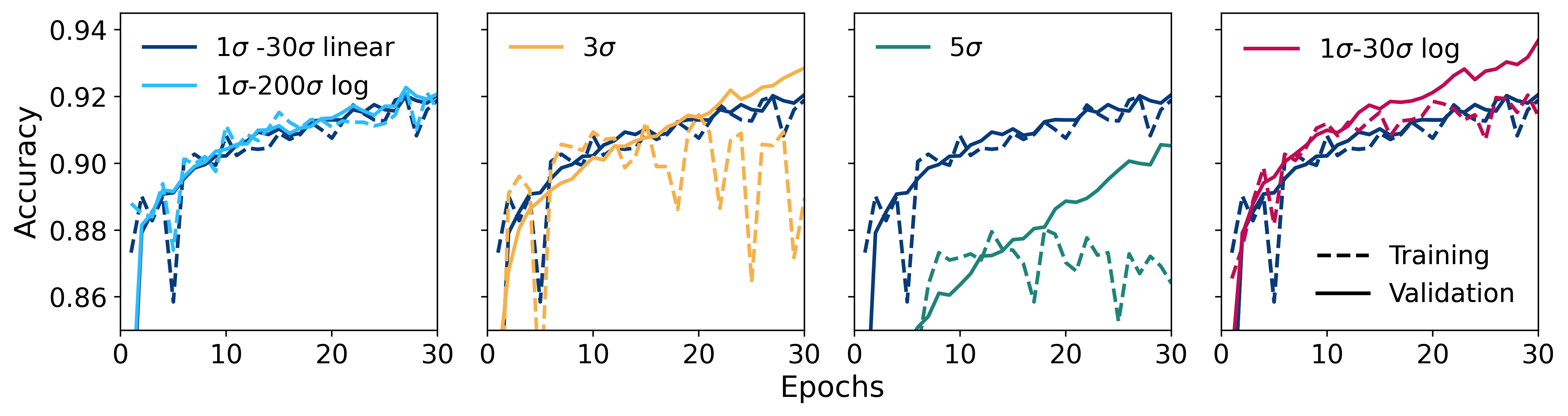}
\includegraphics[scale=0.5]{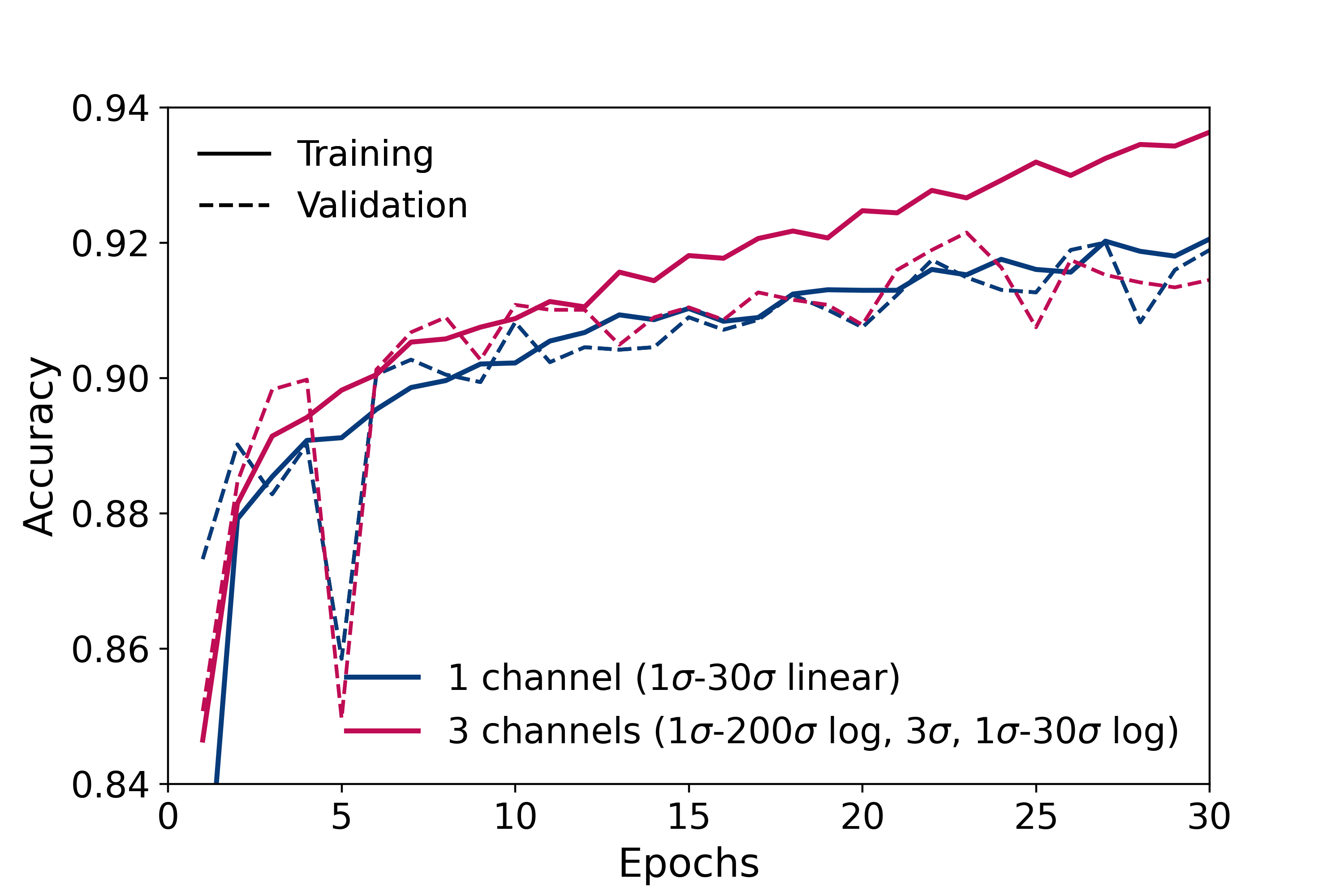}
\end{minipage}
\caption[Baseline sigma-clipping experiments]{Experiments using the baseline architecture with 1 channel and different individual sigma cuts (top row), and the final adopted 3 channels, which combines the 3 individual sigma cuts (bottom).
In each plot the dark blue lines represents the baseline model with 1$\sigma$-30$\sigma$ in linear scale, which is compared to the performance of the model using different sigma cuts, both on the training and on the validation set.}
\label{fig:baseline_sigmacuts1d}
\end{figure*}

Figure~\ref{fig:baseline_sigmacuts1d} compares the performance of the model for different options of the sigma-clipping, on both the training and validation sets. The baseline resulted in a good performance, and the model using the images created using 1$\sigma$–200$\sigma$ logarithmic scale have a very similar performance with a slight improvement, in particular on the validation set. Using the 1$\sigma$–30$\sigma$ stretch in the log scale outperforms the one in the linear scale in both the training and validation sets. However, it requires attention for a higher number of epochs since it tends to overfit after around 20 epochs of training. The 3$\sigma$ cut shows good performance on the validation set but only up to around 15 epochs of training, after which the results start to get unstable. Even though this is the least reliable of the three channels that were ultimately used, it is able to provide some helpful information (as can be seen from the 1-channel network alone). The 5$\sigma$ cut performs poorly in terms of overall accuracy and overfitting, and hence it was excluded. This may be due to a significant loss of information because the majority of extended emission will be below 5$\sigma$ and therefore will be rejected.

The CNN model can be designed to process a three-channel input image. Since the performance of the model is different with different sigma cut images, we can combine the most suitable sigma cuts for the classifier \citep[e.g.][]{Mostert2022}.

The three adopted channels are the 1$\sigma$–200$\sigma$ and 1$\sigma$–30$\sigma$, both in logarithmic scale, and the 3$\sigma$ cut. These were chosen as they were the best-performing individual channels. Each one of them provides subtly different information, and by combining the information from the three channels we provide more details for the training process. The combination of the three images provides an improved performance on the training sample, although the performance on the validation sample is comparable to the 1-channel model. This indicates an increased risk of over-fitting, in particular for higher number of epochs. This aspect will be mitigated later by data augmentation and additional adjustments to the network; we show in Section~\ref{sec:multimodel} that in the final architecture, the 3-channel CNN outperforms the 1-channel version.

\subsection{Multi-modal model}
\label{sec:multimodel}

We created a fusion classifier (a model that can integrate multiple data sources or modalities), by combining the CNN with an artificial neural network (ANN), thus combining images and tabular data into a single multi-modal (MM) architecture.
Each \texttt{PyBDSF} source in the dataset is processed, with radio images being fed into the CNN and features into the ANN.
This approach enables an effective combination of different types of data, thereby further improving the performance of the model.
In our approach, we adopt late fusion (i.e. the process of combining the input data), where the outputs from the CNN and the ANN are concatenated and then passed through two dense fully-connected layers followed by an activation softmax function, generating binary predictions. Other approaches exist, such as early fusion, hybrid fusion (combining early and late fusion), and mid-fusion \citep[e.g. transfer module to fuse CNNs at different stages of the architecture;][]{joze2020mmtm}. There is a debate regarding the impact of fusion techniques on multi-modal model performance, but we do not explore this and focus solely on late fusion in this work.

It should be noted that retrieving the original input from the tabular data features is not feasible since these are only properties based on a combination of Gaussian models; hence, they do not fully describe the original image. However, the tabular features can benefit the multi-modal model by helping to identify characteristics in the images that are more likely to have astrophysical relevance, as well as bringing in information about the multi-wavelength data that goes beyond just the radio images. 

The CNN architecture and hyperparameters used are as defined in Section~\ref{sec:baseline}, with the three-channel input defined in Section~\ref{sec:sigmacuts}. The ANN used for running the experiments is a ANN with two fully connected layers, each with 64 neurons. The model is optimised at later stages, albeit with minimal modifications, as detailed in Section~\ref{sec:performance}.
The initial set of features (baseline features) are the major and minor axes, total and peak flux, and the total number of Gaussians that make up a \texttt{PyBDSF} source, which are the same as those used in the baseline of \cite{alegre2022}. 

Figure~\ref{fig:exp} compares the accuracy, precision and recall of different experiments (see Appendix~\ref{sec:ML_metrics} for a description of these ML performance metrics). As can be seen, the multi-modal model with baseline features shows an increment in the performance values to the CNN alone by about 0.5 per cent in accuracy and about 1 per cent in recall, with negligible effect on precision. It also shows that the performance of recall is superior to that of precision; this is a favourable differential, since the recall (representing the percentage of actual MC sources correctly identified by the model) is the parameter we aim to optimise over precision (which reflects the percentage of sources classified as MC sources by the model that are indeed MC sources). This higher value for recall over precision was already seen in the 3-channel CNN, where the 3-channel CNN had a $\approx 2$ per cent higher recall than the 1-channel CNN, despite a slightly lower overall accuracy.  

\begin{table}
 \caption{Baseline, optical, the set of 18 final features from the \citet{alegre2022} Gradient Boosting Classifier model (GBC), and the full set of 32 features, which include the GBC features and the first 3 nearest neighbours (3 NNs). The features listed in italic are removed (to avoid duplication) when using the 32 features.
 The LR features were scaled using the LoTSS DR1 threshold LR value of $L_{\text{thr}}$ = 0.639. Sources refer to \texttt{PyBDSF} sources, for which the full set of feature values are provided in the online material. The features were computed using the LoTSS DR1 \texttt{PyBDSF} source and Gaussian catalogues, as well as the LR values.$^*$}
 \label{tab:features}
 \begin{tabular}{ll}
  \hline
  Features & Definition \& Origin\\
  \hline
  \textbf{Baseline}\\
  Maj & Source major axis [arcsec]$^a$ \\
  Min &  Source minor axis [arcsec]$^a$\\
  Total$\_$Flux & Source integrated flux density [mJy]$^a$\\
  Peak$\_$Flux & Source peak flux density [mJy/bm]$^a$\\
  log$\_$n$\_$gauss & No. Gaussians that compose a Source$^b$\\
  \hline
  \textbf{Optical}\\
  log$\_$lr$\_$tlv & Log$_{10}$(Source LR value match/$L_{\text{thr}}$)$^c$\\
  lr$\_$dist & Distance to the LR ID match [arcsec]$^c$\\
  log$\_$gauss$\_$lr$\_$tlv &  Log$_{10}$(Gaussian LR/$L_{\text{thr}}$)$^c$\\
  gauss$\_$lr$\_$dist & Distance to the LR ID match [arcsec]$^c$ \\
  log$\_$highest$\_$lr$\_$tlv & Log$_{10}$(Source or Gaussian LR/$L_{\text{thr}}$)$^c$\\
  \textit{log$\_$NN$\_$lr$\_$tlv}& Log$_{10}$(LR value of the NN/$L_{\text{thr}}$)$^c$\\
  \textit{NN$\_$lr$\_$dist} & Distance to the LR ID match [arcsec]$^c$\\
  \hline
  \textbf{GBC (baseline \& optical)}\\
  gauss$\_$maj & Gaussian major axis [arcsec]$^b$\\
  gauss$\_$min & Gaussian minor axis [arcsec] $^b$\\
  gauss$\_$flux$\_$ratio & Gaussian/Source flux ratio$^{a,b}$\\
  NN$\_$45 & No. of sources within 45\arcsec$^a$ \\
  \textit{NN$\_$dist} & Distance to the NN [arcsec]$^a$ \\
  \textit{NN$\_$flux$\_$ratio} & NN flux/Source flux density ratio$^a$\\
  \hline 
  \textbf{Nearest Neighbour (3 NNs)}\\
  \textbf{(All feat. replacing italic ones)} & \\
  NN$\_$Maj (x3) & NNs major axis [arcsec]$^a$ \\
  NN$\_$Min (x3) &  NNs minor axis [arcsec]$^a$\\
  NN$\_$log$\_$lr$\_$tlv (x3) & Log$_{10}$(LR value match/$L_{\text{thr}}$)$^c$\\
  NN$\_$lr$\_$dist (x3) & Distance to the LR ID match [arcsec]$^c$\\
  NN$\_$dist (x3) & Distance to the NNs [arcsec]$^a$ \\
  NN$\_$flux$\_$ratio (x3) & NNs flux/Source flux density ratio$^a$ \\
  \hline
 \end{tabular}
  \parbox{\columnwidth}{$^*$ a - \texttt{PyBDSF} radio source catalogue \citep{Shimwell2019lofar};\\
  b - \texttt{PyBDSF} Gaussian component catalogue \citep{Shimwell2019lofar};\\
  c - Gaussian and \texttt{PyBDSF} LR catalogues \citep{williams2019lofar};\\}
\end{table}

\begin{figure*}
\begin{minipage}{\textwidth}
\centering
\includegraphics[width=0.7\columnwidth]{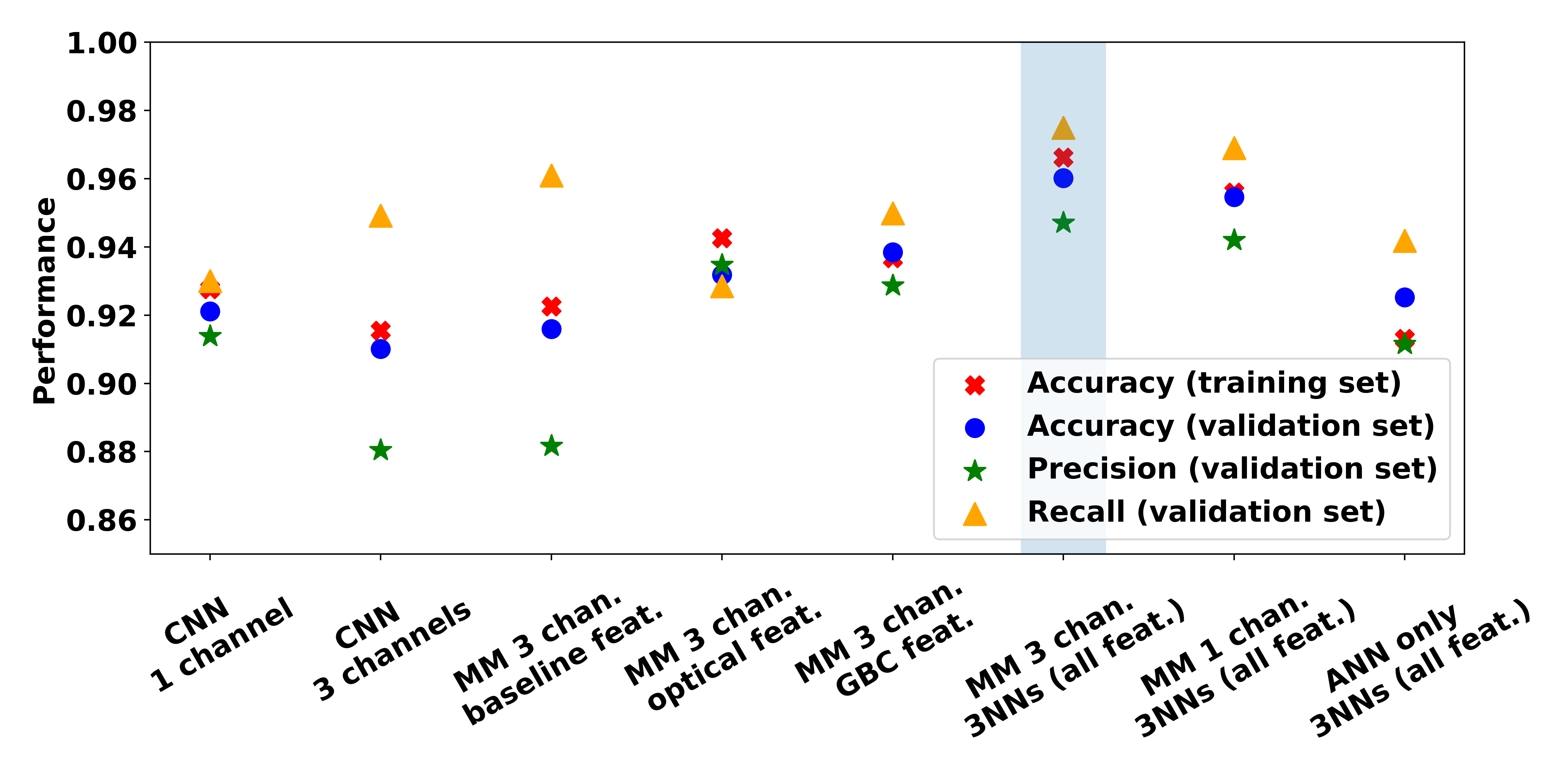}
\end{minipage}
\caption[Experiments]{Main set of experiments, with values for accuracy on the training (red x) and validation (blue circles) sets, with precision (green stars) and recall (yellow triangles) for class MC also shown for the validation set; in each case, the plotted points correspond to different epochs where the training and validation sets show the best performance possible for similar results on both sets (i.e. training was stopped before significant signs of overfitting). The F1-score (not displayed) is consistent with the accuracy values within 10$^{-4}$. The CNN 1 channel corresponds to the baseline model, for which the performance for all the metrics on the training and validation sets is very similar. The introduction of a 3-channel CNN (CNN 3 channels) and the modification of the architecture into a multi-modal (MM) model (MM 3 chan. baseline feat.) helped to greatly increase the recall. The introduction of more features, independently, all helped to improve the performance. Plotted points correspond to the baseline features (MM 3 chan. baseline feat.), only the optical features (MM 3 chan. optical feat.), and the 18 features used in the GBC model (MM 3 chan. GBC feat.) from \cite{alegre2022}. The best results were obtained from combining the 18 GBC features with additional information about the first, second and third NNs (MM 3 chan. 3NNs all feat.) shown as the shaded model. Overall, all the metrics improved from around 92 per cent to 96 per cent as a result of adopting a MM model and adding more features. For comparison, also shown are the MM model using only 1 channel (MM 1 chan. 3NNs, all feat.), and the neural network alone (ANN only 3NNs, all feat.), both of them showing inferior performance.}
\label{fig:exp}
\end{figure*}

Different sets of features were then tested independently, building upon the features developed by \citet{alegre2022}. See Table~\ref{tab:features} for details of the different features and Figure~\ref{fig:exp} for a summary of the experiments. The features denoted with `lr' on Table~\ref{tab:features} are based on the Likelihood Ratio (LR) values derived from \citet{williams2019lofar} and correspond to the likelihood of a LoTSS radio source having a true optical galaxy counterpart (Pan-STARRS, if available, or otherwise infrared WISE sources). The LR is a statistical technique that has long been used to automatically cross-match sources at different wavelengths \citep[e.g.][]{Sutherland1992}, in particular those with longer wavelengths for which the positional uncertainty is greater due to the large beam size of the telescopes, resulting in multiple possible counterparts. The LR assesses the probability of a galaxy having a true radio counterpart based on the positional uncertainty of radio sources and both the magnitude distributions of the true counterparts and the source counts of the background sources.

First, we considered only optical features; these comprise the log of the LR relative to the threshold value (tlv; that is, the LR divided by the lowest LR value at which a cross-match is considered to be genuine) and the distance to the highest LR counterpart, for both the source, the first nearest neighbour (NN), and the Gaussian with the highest LR value, as well as the highest log LR tlv between the source and the Gaussian. Using only optical features resulted in an increase in precision, a decrease in recall, and an overall increase in accuracy to about 93 per cent and 95 per cent accuracy on the validation and training sets, respectively. Second, using the set of 18 final features defined in \cite{alegre2022} improves the recall and leads to similar accuracy values of 94 per cent on both training and validation sets. 
The NNs have been shown to improve the model of \cite{alegre2022}, as they provide useful information about the source surroundings. Therefore, we expand this to incorporate additional NNs, in particular the second and third NNs; for each one, the set of features includes the minor and major axes, the log of the LR tlv, the LR distance, the distance to the NNs and the flux ratio between the NNs and the source. 
Adding this information about the second and third NNs to the previous 18 features proved to significantly improve all the metrics by almost 2 per cent each (see Figure~\ref{fig:exp}). Experiments using more NNs, such as including the fourth and fifth, did not reveal any further improvements.
The results show that the NNs feature information is essential to identifying MC sources since it leads not only to better overall model performance but also to higher values of recall.
Additional experiments on features, such as feature scaling or replacing measured axis sizes with their deconvolved equivalents, failed to produce any further improvement to the model (or decreased performance) and so are not considered further.
As a final test of the performance of our multi-modal model, we show on Figure~\ref{fig:exp} also the performance of the model with the full set of features, but including only 1 channel of input image for the CNN (the baseline 1-30$\sigma$ cut). This shows the performance in all metrics drops by about 1 per cent compared to the 3-channel CNN, justifying our decision to use the 3-channel model. We also show the performance of the ANN alone (i.e. without the CNN). Like the CNN alone, this achieves an overall accuracy on the validation set of around 92 per cent, considerably below that of the multi-modal model.

\subsection{Removal of Small Isolated Single Gaussian (SISG) sources}
\label{sec:nsisg}

In this section, a particular set of sources (hereafter referred to as SISG sources) is removed from the dataset in order to evaluate whether it results in any improvements in the performance of the classifier. These correspond to small (major axis smaller than 15 arcsec), isolated (no NNs within a 45 arcsec radius), and single Gaussian \texttt{PyBDSF} sources, which were not cross-matched with a large optical ID. The SISG sources correspond to a large proportion of the sources (186,371 \texttt{PyBDSF} sources, or 57.7 per cent of the full LoTSS DR1 sample, excluding artefacts), for which the classifier from \cite{alegre2022} achieved 99.98 per cent accuracy. 
The vast majority of the sources in this group can be cross-matched using the LR method. It is characterised by single-component sources, with the exception of 133 components (the cores) of MC sources, and 4 additional single-component sources for which the LR method gave an incorrect ID. 
This group of sources shows broadly uniform properties, and they do not add diversified information about class S. Excluding these objects from the training sample therefore allows the classifier to be exposed to a wider range of class S sources. 

The SISG sources are the type of sources that can be processed even without a classifier since, by their characteristics, they can be cross-matched by LR methods. Additionally, they are also the ones the classifier would easily identify as single-component sources and therefore are likely to get the correct classification. The SISG sources were, therefore, removed from training and testing on a balanced dataset, and they are assigned automatically to class S if they are presented in the data.~\footnote{The SISG sources are indicated in the table provided as complementary online material.} 

When removing the SISG from the training set, the model drops in performance on the training set (about 1 per cent to 1.5 per cent worse ability to distinguish between the classes). 
The overall decrease in performance can be attributed to the exclusion of the easily-classifiable 60 per cent of the single-component sources.
With SISG removed from class S, this class is now characterised by sources that are more complex (i.e. this class has a higher number of sources that are not isolated, clustered, and composed by multiple Gaussians) and therefore the performance in class S drops. But at the same time class S now comprises elements that are more relevant for the classification. 
Importantly, if the model is applied to the full imbalanced dataset, it performs better (especially on class S) than the model trained on all sources (see Section~\ref{sec:fulldataset}). 
These results indicate that exclusion of the SISG sources improves the overall performance on the full dataset. This strategy also reduces computational costs since it eliminates the need to process more than 50 per cent of the data, which is particularly important when processing large samples.

\subsection{Augmentation}
\label{sec:augmentation}

When removing the set of SISG sources from the dataset, the model tends to overfit. In order to minimise this issue, we use data augmentation by increasing the number of examples of the minority class. Data augmentation is an artificial way of enlarging the training set by creating alternate samples of the original data. A common approach to achieve this is by generating synthetic examples, typically through the application of geometric or colour transformations \citep[see][for a review]{shorten2019survey}. This technique is commonly used in deep learning models, as it is often necessary to avoid overfitting since these models require a higher number of examples to be trained \citep[e.g.][]{goodfellow2016deep}. 
In astronomy, \cite{dieleman2015rotation} applied augmentation by using geometric transformations to prevent a CNN model from learning specific orientations of galaxies in optical images. Assuring the models are rotational invariant is now a common practice for astronomy applications (see also Appendix~\ref{sec:rotation}). In radio morphology classification, where there are generally 2–5 classes but sometimes as few as 100 objects per class \citep[e.g.][]{Aniyan2017}, augmentation is commonly achieved by massive oversampling, e.g.\ applying multiple rotational and flipping angles. \cite{Viera2021} demonstrated that the use of both vertical and horizontal flips increased accuracy by roughly 10 per cent, but improper augmentation operations, such as shifting and zooming, degraded their CNN model.

\begin{figure}
\centering
\vspace{0.5cm}
\includegraphics[width= \columnwidth]{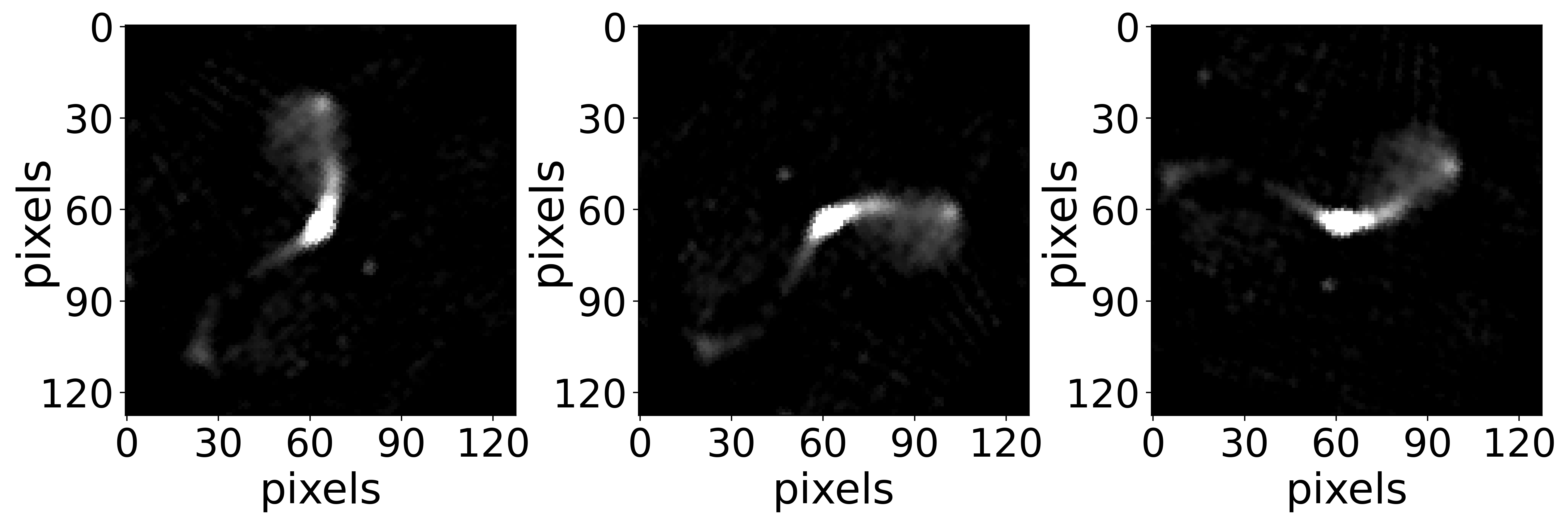}
\caption[Augmentation process example]{Example of the augmentation process, where sources undergo random rotations and horizontal and vertical flips. This is done after sigma clipping (1$\sigma$-30$\sigma$ linear in this example) on 256$\times$256 pixel images before cropping them to 128$\times$128 pixels.}
\label{fig:augmentation example}
\end{figure}

The augmentation procedure was done as follows: having cut 256$\times$256 pixel FITS images from the original LoTSS DR1 mosaics and applied different sigma clipping thresholds, we then performed augmentation on the minority class (class MC). We rotated each image around the \texttt{PyBDSF} position at the centre of the frame by a random angle between 0 and 2$\pi$ and applied random (true or false) vertical and/or horizontal flipping. The transformed images were then cut to their final sizes of 128$\times$128 pixels (see Figure~\ref{fig:augmentation example} for an example). 
By rotating the images prior to reducing their size, we avoid the issue of empty corners created by the rotation; this avoids the need for any interpolation to fill in the empty regions and eliminates the possibility of the classifier correlating such corner effects to the augmented class.

The majority of the sources in LoTSS correspond to sources belonging to class S. The training set for class S was created by randomly undersampling single-component sources and therefore did not require any type of augmentation. The blended sources, which are rare, were also not augmented. They were added up to the undersampled single-component sources in order to ensure the same number of sources as in class MC. This allows for the creation of a balanced dataset for evaluating the results. Even though balanced datasets are not typical, balancing the dataset is necessary for the network to effectively learn the characteristics of the sources in the different classes. 

\begin{figure}
\centering
\includegraphics[width=\columnwidth]{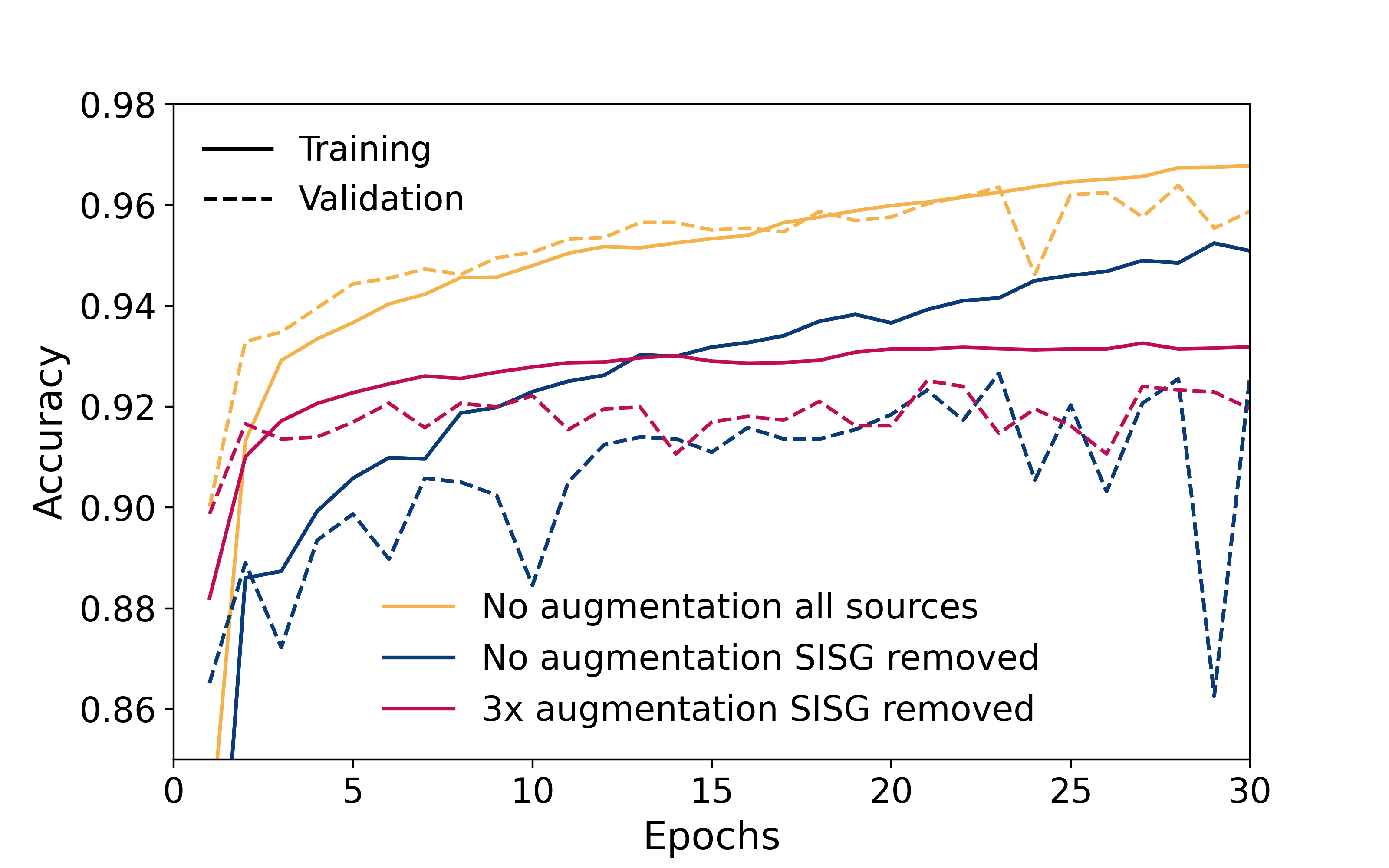}
\caption[Augmentation effect on the dataset without SISG sources]{The figure shows the effect of augmentation when training with the SISG sources removed from the dataset. The coloured lines represent the learning curves for 30 epochs of training for illustration. The model trained with all types of sources achieves greater accuracy (yellow line), but this is also because the class S contains about 60 per cent of sources that are easy to classify; see the main text for a discussion. The performance drops and shows major discrepancies on the training and validation sets when the SISG sources are removed (blue line), but augmentation helps to compensate for this effect (red line).}
\label{fig:augmentation nsisg}
\end{figure}

The augmentation process is exclusively applied to the training set and only to the minority class, as mentioned previously. We experimented with increasing the number of sources in the dataset by factors of two, three, and five relative to the original dataset. The validation and test sets remained unaffected and contained always the same amount of sources, regardless of augmentation. The datasets were constructed using the same sources, but for each augmentation factor, new single-component sources from the majority class were added. When using the augmented datasets, it was necessary to adjust the learning rate. We found that augmenting the dataset three times the original size was sufficient to prevent overfitting while achieving good results, as we can see from Figure~\ref{fig:augmentation nsisg}.
On three times the size of training set, the number of sources in class MC is 18,789 MC sources, which is three times the number of MC sources in the original dataset (excluding any MC source for which at least one the source components was in the SISG group). The number of sources in class S is also 18,789, but in this case, these correspond to 18,189 single-component sources and 600 blended sources.~\footnote{Information about the dataset splitting is provided in the online table. The following values are assigned to sources associated with each set of data (as indicated by the column \textit{mc$\_$}dl$\_$dataset): none of the sets (0); training set (1); test set (2); validation set (3).}

In the context of the multi-modal model, it was necessary to replicate the feature values for every augmented image, ensuring that they align with each respective instance. Furthermore, we also ensured that the dataset was properly shuffled when training the classifier.

\subsection{Model optimisation and final architecture adopted}
\label{sec:optimisation}

In order to investigate if it is possible to optimise the model, different aspects were analysed. These comprise architectural variations and hyperparameter adjustments. Changes to the structure included investigating the ANN width and depth (by varying the number of layers and changing the number of neurons in each layer), the removal of layers in the CNN, and the presence or absence of layer normalisation or batch normalisation.
Adding layers to the ANN part of the model, specifically ranging from 2 to 5 layers, did not result in improvements. On the other hand, changing the number of hidden neurons in each layer (64, 128, 512, and 256) showed that using 256 neurons resulted in higher performance (with a further reduction in the dropout rate from 50 percent to 25 percent). As a result, the ANN part of the model adopted is a two-layer ANN containing 256 neurons each.
Regarding the CNN module, the use of batch normalisation following each convolutional or dense layer, either as a substitute or in combination with dropout, led to a decline in performance. Furthermore, we investigated reducing the length of the CNN. However, it was observed that the elimination of the first layer led to overfitting and a decrease in overall performance.

\begin{figure}
\centering
\includegraphics[scale=0.4]{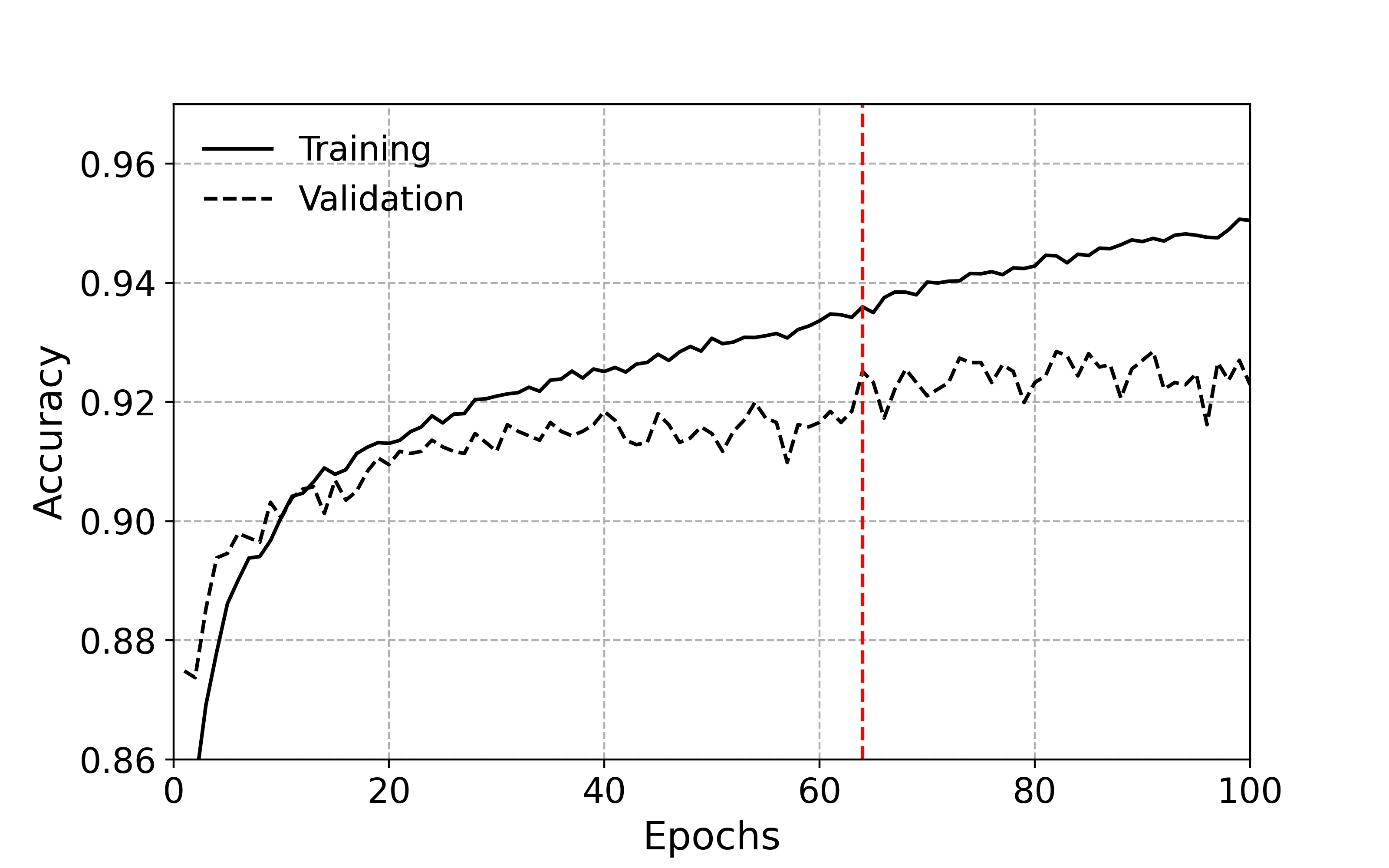}
\caption[Learning curve for the final adopted model] {Learning curve for the final adopted model after optimisation. 
The model reaches about 91 per cent accuracy after only 20 epochs of training on both the training and validation sets. 
Training for longer gives about 1 per cent improvement to 92 per cent accuracy on the validation set. There is a higher difference in the performance on the training set with an increasing number of epochs, which is a clear indication that the model may be overfitting. However, as we can see from Table~\ref{tab:statistics_dl_datasetA}, it is worth training for longer since the performance on both validation and test sets ends up being very similar, and so training for longer helps improve the model by about 1 per cent in accuracy. We defined epoch 64, which was selected from the 60–70 range of epochs where the performance seems to stabilise. The accuracy reaches a plateau on the validation set and does not seem to improve more than about 92 per cent.
}
\label{Fig:OPTmodel}
\end{figure}

Furthermore, we explored alternative hyperparameters besides the baseline ones defined in Section~\ref{sec:baseline}. One of the experiments involved testing different learning rates, including ones with both static and variable rates, and using alternative learning optimisers which iteratively adjust the weights of the networks and/or the learning rate of the training to find the minimum error for a certain problem.
Using the Adam optimiser \citep{kingma2014adam} yielded inferior performance, while using stochastic gradient descent \citep[SGD; e.g.][]{bottou2010large}, particularly when used with momentum, demonstrated superior performance. The best results were achieved with a SGD, which is an optimiser that adapts the weights but not the learning rate. The weights were updated using Nestrov momentum \citep{sutskever2013importance}. Different batch sizes were evaluated since smaller batch sizes tend to result in higher performance, although the extent of their effectiveness depends on the GPU being used since very small batch sizes may cause memory problems. Different values of batch sizes were assessed, including 16, 32, and 128. Results were indeed better for smaller batches, and 32 was chosen as the best without massive computational problems.
Additionally, the optimisation process involved reviewing the number of training epochs and eventual early stopping. Training the model for a higher number of epochs (more than 50) resulted in accuracy in the validation set above 92 per cent, with no significant differences in performance on the training set, as can be seen from Figure~\ref{Fig:OPTmodel} . We identified an interval of 10 epochs, ranging from the 60\textsuperscript{th} to the 70\textsuperscript{th} epoch, which led to the most favourable results. These epochs show strong performance and smaller overfitting, with a discrepancy between the validation and training sets of less than 1.5 per cent. It was also observed that training below this range leads to a decline in performance, with accuracy dropping below 92 per cent. The chosen epoch for stopping training was epoch 64, because this results in only a minor difference of 1.083 per cent between the training and validation sets. This results in a training accuracy of 93.6 per cent and a validation accuracy of 92.5 per cent.

\begin{figure*}
\begin{minipage}{\textwidth}
\centering
\includegraphics[width = 0.8 \columnwidth]{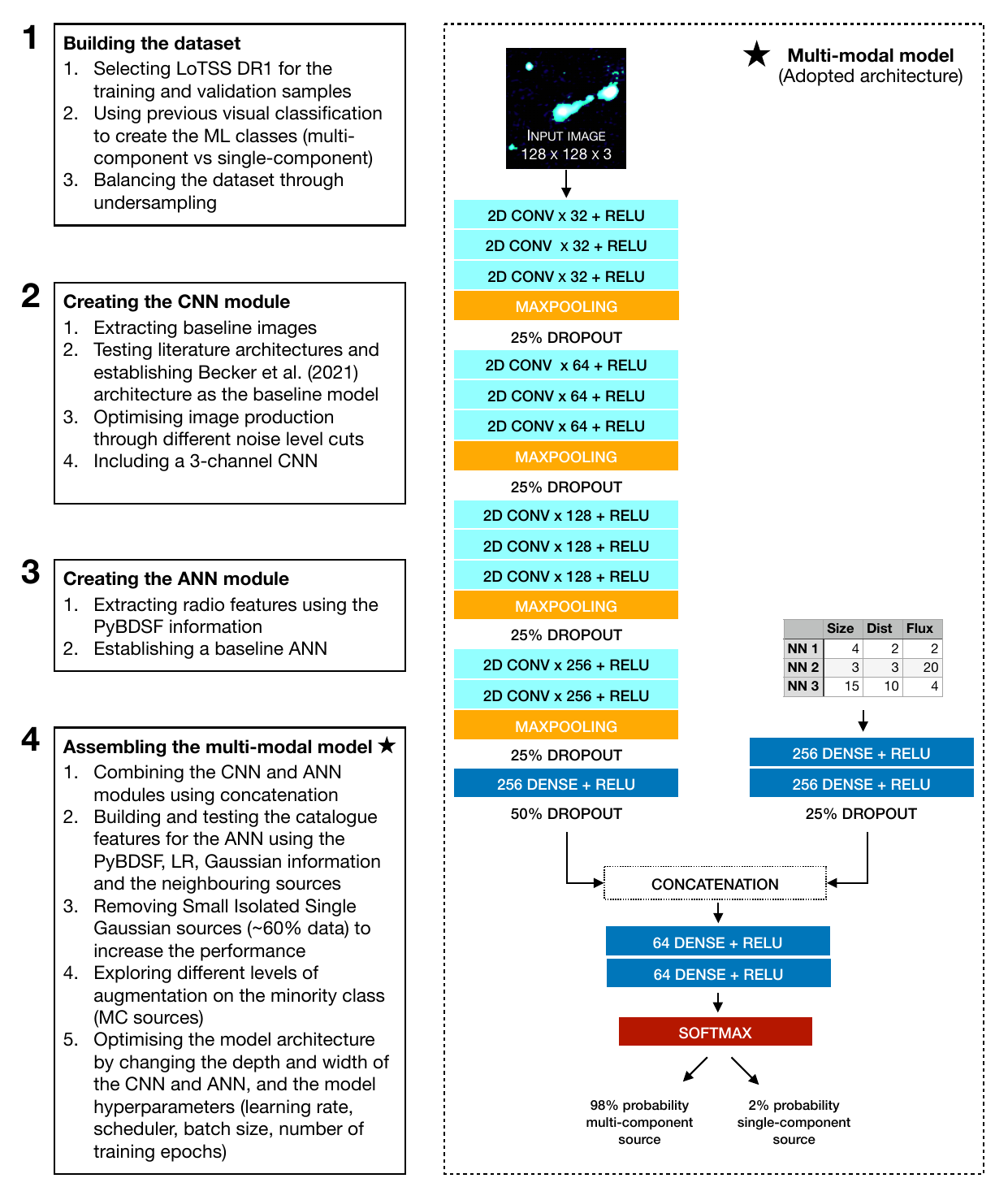}
\caption[Multi-modal architecture for the optimised model]{Sequence of steps employed to construct the final model (left), and the adopted model  architecture (right). It consists of a multi-modal architecture, that inputs a 3-channel image of 128$\times$128 pixels into a 4-block CNN (very similar to the \cite{becker2021cnn} architecture), and a set of features into a 2-layer ANN. The model outputs the probability of a source being a multi-component (class MC) or a single-component source (class S). More details about the architecture and the model hyperparameters can be found in the main text.} 
\label{fig:model_adopted}
\end{minipage}
\end{figure*}

Figure~\ref{fig:model_adopted} provides a schematic representation of the adopted architecture and outlines the steps taken to achieve the final model. These comprise 1) building the dataset, 2) creating the CNN, 3) and the ANN modules, and 4) assembling the multi-modal model (including optimisation). The model inputs a 3-channel radio image into a 4-block CNN and a set of features into a 2-layer ANN with 256 neurons each. Each convolutional layer has a kernel of 3x3, padding of 1, and stride of 1 (with the exception of the first layer of the first two blocks of the CNN, which have a stride of 2), followed by a ReLU activation function. The maxpooling layer has a kernel size of 2x2, a stride of 1, and padding of 1. The outputs of the CNN are then concatenated with the outputs of the ANN and passed through a set of two dense layers with 64 neurons each before being fed into a softmax function, which outputs a probability of the source being a MC source or not. The model was trained for 64 epochs with a batch size of 32, a SGD optimiser with a 0.9 Nestrov momentum, and a learning rate of 0.0001 without decay. The number of filters in the convolutional layer is indicated in the figure, as is the amount of dropout applied.

\subsection{Final model performance}
\label{sec:performance}

Performance metrics using the optimised model trained on the augmented, balanced dataset are presented in Table~\ref{tab:statistics_dl_datasetA}, for both the validation and test sets. The value adopted for the threshold is 0.5, which is commonly used for balanced datasets, and the metrics used are accuracy, precision, recall, and F1-score, as explained in Appendix~\ref{sec:ML_metrics}.
Given that the dataset adopted for training the model was created with the SISG removed, the results presented here are for a dataset where the SISG were removed as well. As can be seen from the table, the performance on the validation and test sets is very similar across all of the metrics, which shows the model is able to generalise to unseen data.

Overall, the model favours recall on class MC (and precision on class S), which is the value we want to optimise. Our goal is to maximise the number of correctly identified MC sources because it will ensure accurate source flux measurements. If these are sent to be cross-matched automatically without prior analysis, the source properties will be wrong. At the same time, we want to keep the number of sources wrongly identified as MC sources low, either because the source component association algorithm may fail on those and/or because we will have to analyse those sources and manually grouping them and/or cross-matching.\\

According to the values obtained for the recall on the validation and test sets, the model is able to identify 94 per cent of the sources that are MC sources correctly. From the ones that are classified as not being MC sources, about 94 per cent as well are indeed not MC sources, as per the precision obtained in class S for both validation and test sets. 

Despite the increased number of complex sources in the augmented dataset (which is accompanied by the same number of single-component sources), the classifier effectively differentiates between the various classes.
This shows the ability of the classifier to handle rotation invariance since about 60 per cent of the sources in class MC suffered rotations and flippings. More details of testing the final model to ensure it has rotation/reflection symmetry, confirming that it does, are discussed in Appendix~\ref{sec:rotation}.

\begin{table}
 \caption[Multi-modal model performance on a balanced dataset]{Performance on a balanced dataset for the final model with SISG sources removed. The validation and test sets each contain 2,685 and 2,683 sources, respectively, with an equal distribution of sources between class MC and class S as defined in Section~\ref{sec:classes}. The results show the accuracy, precision, recall, and F1-score for the 2 classes for a decision threshold of 0.5.}
 \begin{center}
 \begin{tabular}{l c c }
    & Validation set & Test set \\
  \hline
  Accuracy     & 0.925 & 0.925 \\
  F1-score MC  & 0.926 & 0.926 \\
  F1-score S   & 0.924 & 0.923 \\
  Precision MC & 0.914 & 0.911 \\
  Precision S  & 0.937 & 0.939 \\
  Recall MC    & 0.939 & 0.941 \\
  Recall S     & 0.911 & 0.908 \\
  \hline
 \end{tabular}
 \label{tab:statistics_dl_datasetA}
 \end{center}
\end{table}

\section{Application to the full LoTSS-DR1 dataset}
\label{sec:fulldataset}

In this section we apply the model to the full LoTSS DR1 dataset. The LoTSS datasets differ from the data used to train and test the model both in terms of class balance and the type of sources that make up the classes, since the SISG sources were removed from training. Class imbalance happens when one of the classes is severely underrepresented, which is the case for MC sources in the real LoTSS datasets. The classes defined are highly imbalanced, with less than 3 per cent of the sources being MC sources. This effect is commonly counteracted with threshold moving, which can be done by evaluating the metrics we intend to improve and choosing a more suitable threshold value. However, it can be observed that the use of a training set where the SISG sources are removed already goes some way towards counterbalancing the class asymmetry, and with our desire to maximise recall on the MC class, suitable thresholds are found to be around 0.5, as discussed next, which is the default threshold value for balanced datasets.

\subsection{Performance as a function of the threshold}
\label{sec:performance_threshold}

In order to investigate if 0.5 is the appropriate value to discriminate between the classes, we examined the performance of the model on the LoTSS DR1 sample using different threshold values. As outlined in more detail by \citet{alegre2022}, corrections are applied in cases where at least one of the source components is flagged as being a MC source: in these cases, although other components of the same MC source may not themselves be identified as MC (and hence incorrectly classified as false negatives~\footnote{A false negative (FN) source is a class MC \texttt{PyBDSF} source classified by the model incorrectly as class S. A false positive (FP) is a source that is incorrectly classified as a MC source but is actually a class S one (either a single-component source or a blended detection). True positives (TP) and true negatives (TN) are sources that the model has correctly identified, corresponding to class MC and class S sources, respectively.}), these components will be re-found as part of the examination of the identified MC component. To account for this, following \citet{alegre2022}, we remove these sources from the false negative (FN) category.

\begin{figure*}
\begin{minipage}{\textwidth}
\centering
\begin{tabular}{ll}
\includegraphics[scale=0.4]{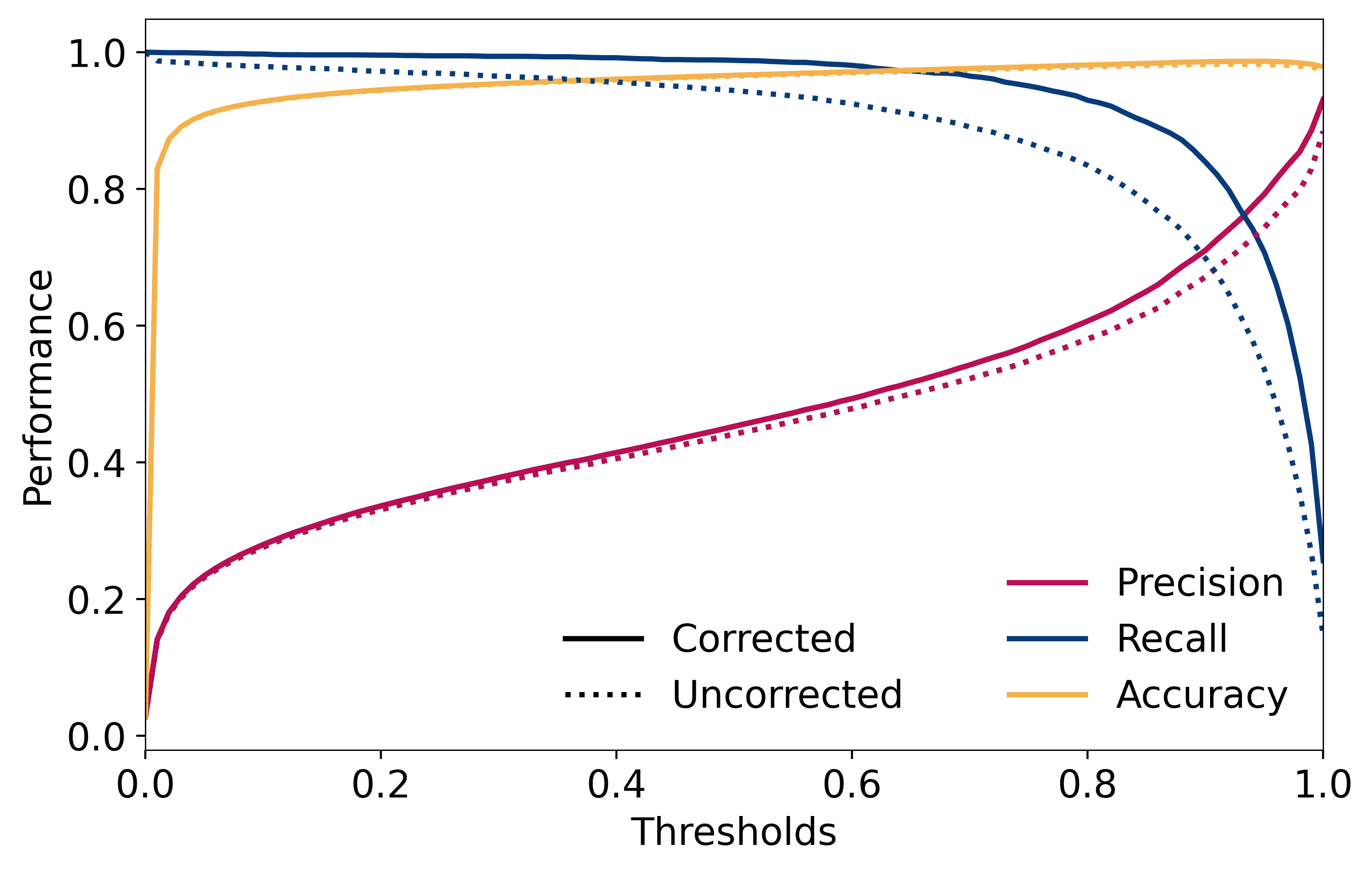}
&
\includegraphics[scale=0.4]{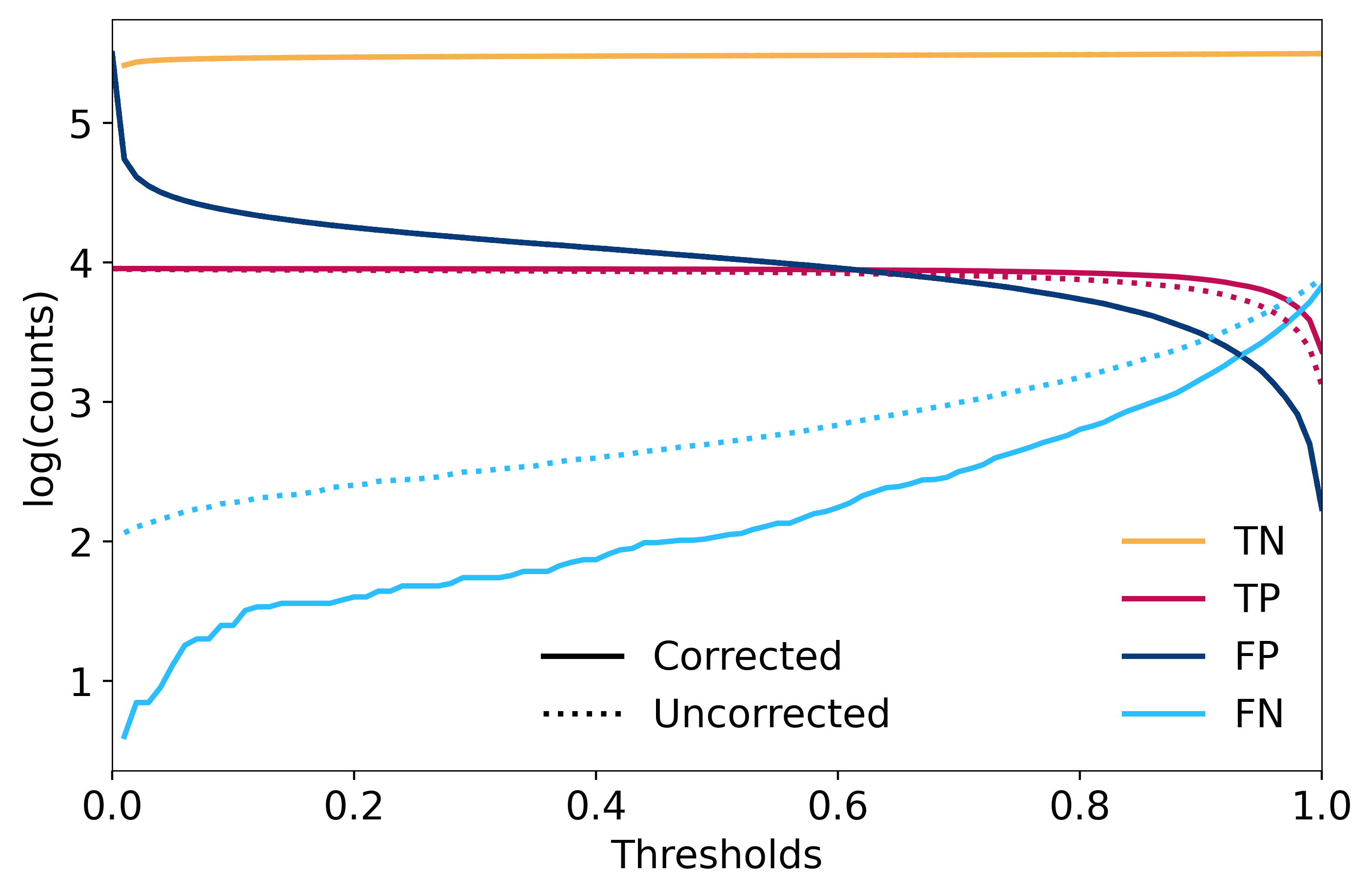}
\end{tabular}
\end{minipage}
\caption[Model performance on LoTSS DR1 in function of the threshold]{Performance of the model on the LoTSS DR1 sample plotted for different threshold values with corrections (solid line) and without corrections (dashed line). Corrections are applied when one source component is identified as part of an MC source allowing the other source components in the same MC source to be recovered, even if they themselves are false negatives. Please see the text for a more detailed explanation. Left: accuracy, recall, and precision. Right: true negative (TN), true positive (TP), false positive (FP), and false negative (FN) counts on a logarithmic scale. The results correspond to the model applied to the full dataset, where sources in the SISG category got assigned automatically to class S, i.e. not being part of a MC source.}
\label{Fig:Race}
\end{figure*}

Figure~\ref{Fig:Race} shows the results of applying the final model to the LoTSS dataset. It can be seen that, in general, the model favours recall instead of precision unless the threshold is above 0.9. Recall is the metric it was intended to prioritise and for which the results show always high values close to unity up to a threshold of 0.6. For thresholds around 0.5 the number of FN reaches values around 100, and it is higher for higher thresholds, reaching values close to 200 at a threshold value of 0.62, which is where the number of TP and FP sources is counterbalanced. A 0.5 threshold shows a good performance for recall and does not compromise precision too much, so this is the threshold value adopted. Depending on the choice of the metrics one intends to optimise, a sensible value range would be between about 0.5 and 0.6 in order to reduce the number of FP, since the true positive rate (TPR) decreases towards higher thresholds, as it will be discussed next.

\begin{figure}
\centering
\includegraphics[width=\columnwidth]{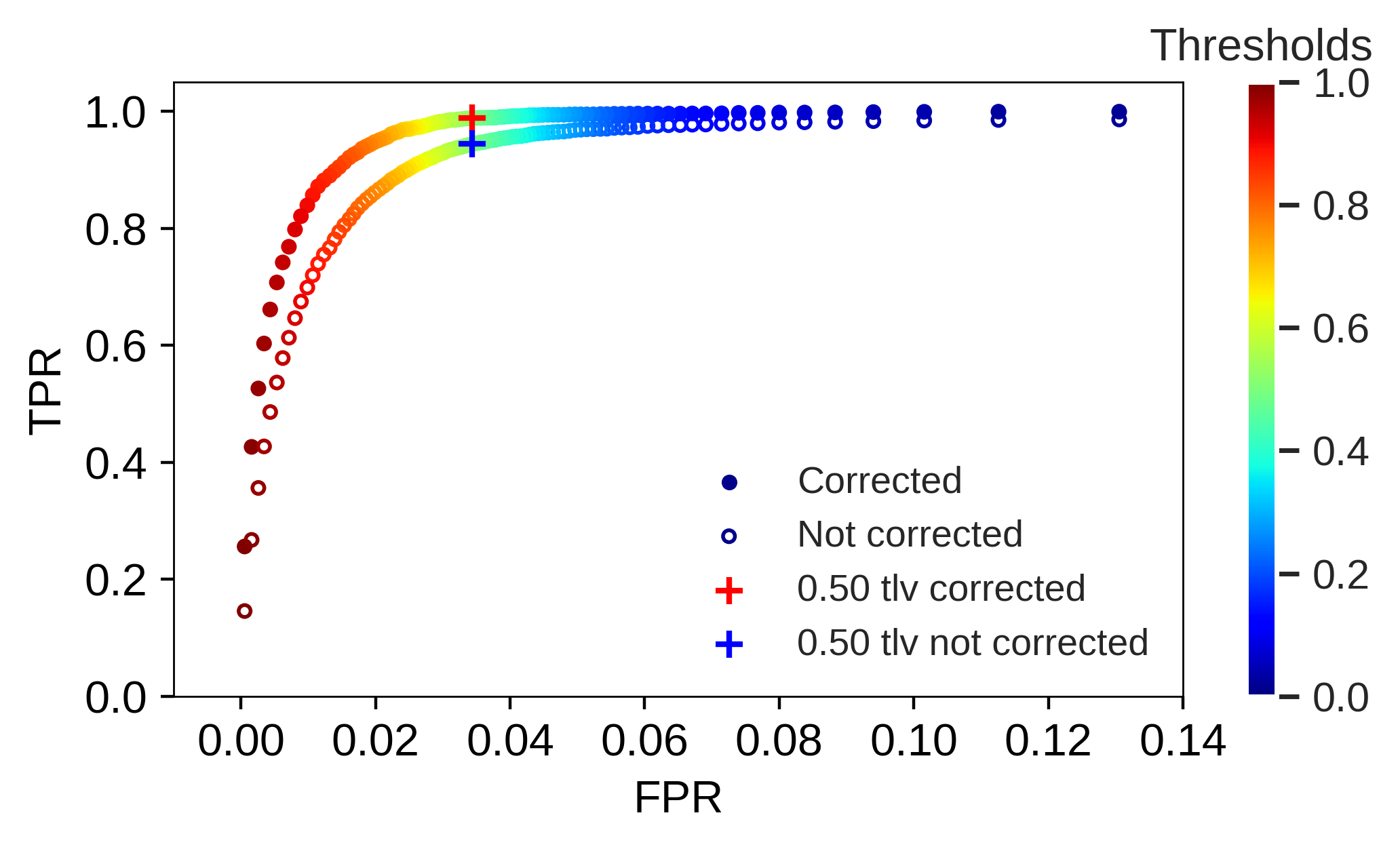}
\caption[ ROC curve on LoTSS DR1]{Receiver operating characteristic (ROC) curve, with the FPR (FP/(FP+TN)) in the x-axis and the TPR (TP/(TP+FN)) in the y-axis, plotted for different threshold values (colour coded). These correspond to values for which corrections are applied (filled markers) or not applied (empty markers). Overall, the classifier shows outstanding performance, and corrections improve the model for both TPR and FPR. Note that the plot corresponds to a zoom in to the left-hand side of the ROC curve, with only relevant values of the axes shown. The cross markers correspond to the 0.5 threshold adopted.} 
\label{fig:roc_full_ds_dl}
\end{figure}

In Figure~\ref{fig:roc_full_ds_dl}, we show the Receiver Operating Characteristic (ROC) curve where the FPR corresponds to the proportion of class S sources that are incorrectly classified as being MC sources, and the TPR corresponds to the proportion of MC sources that are correctly identified by the model (see Appendix~\ref{sec:ML_metrics} for performance metrics).

The FPR values are always very low, but this is because there are many single-component sources in the dataset and therefore many sources that are TN. The adoption of a threshold value of 0.5 (blue and red crosses in Figure~\ref{fig:roc_full_ds_dl}) leads to a FPR of nearly 4 percent, corresponding to approximately 10,000 sources of class S. Only for higher threshold values does the number of FP start to decrease (which can be seen in Figure~\ref{Fig:Race}), and therefore the FPR decreases. This shows that only for thresholds above about 0.8 there is a significant reduction in the number of FP sources and in the FPR.

On the other hand, the TPR values are always very high, decreasing only towards greater thresholds. This is because the number of TP is roughly constant across thresholds (see Figure~\ref{Fig:Race}) decreasing only for threshold values close to unity, and the number of FN is always low in comparison. However, the FN counts start to increase for higher thresholds, and therefore the TPR decreases. For the 50 per cent threshold adopted, this means that almost all the MC sources are being accurately identified, with only a very small number of MC sources being missed by the model (see also Figure~\ref{Fig:CM_full_ds_individuals}).

\subsection{Results at a threshold of 0.5}

\begin{figure*}
\begin{minipage}{\textwidth}
\centering
\begin{tabular}{lll}
\hspace{1cm} SISG sources set as S & \hspace{0.9cm} SISG sources removed & \hspace{1.2cm} SISG sources only \\
\includegraphics[scale=0.33]{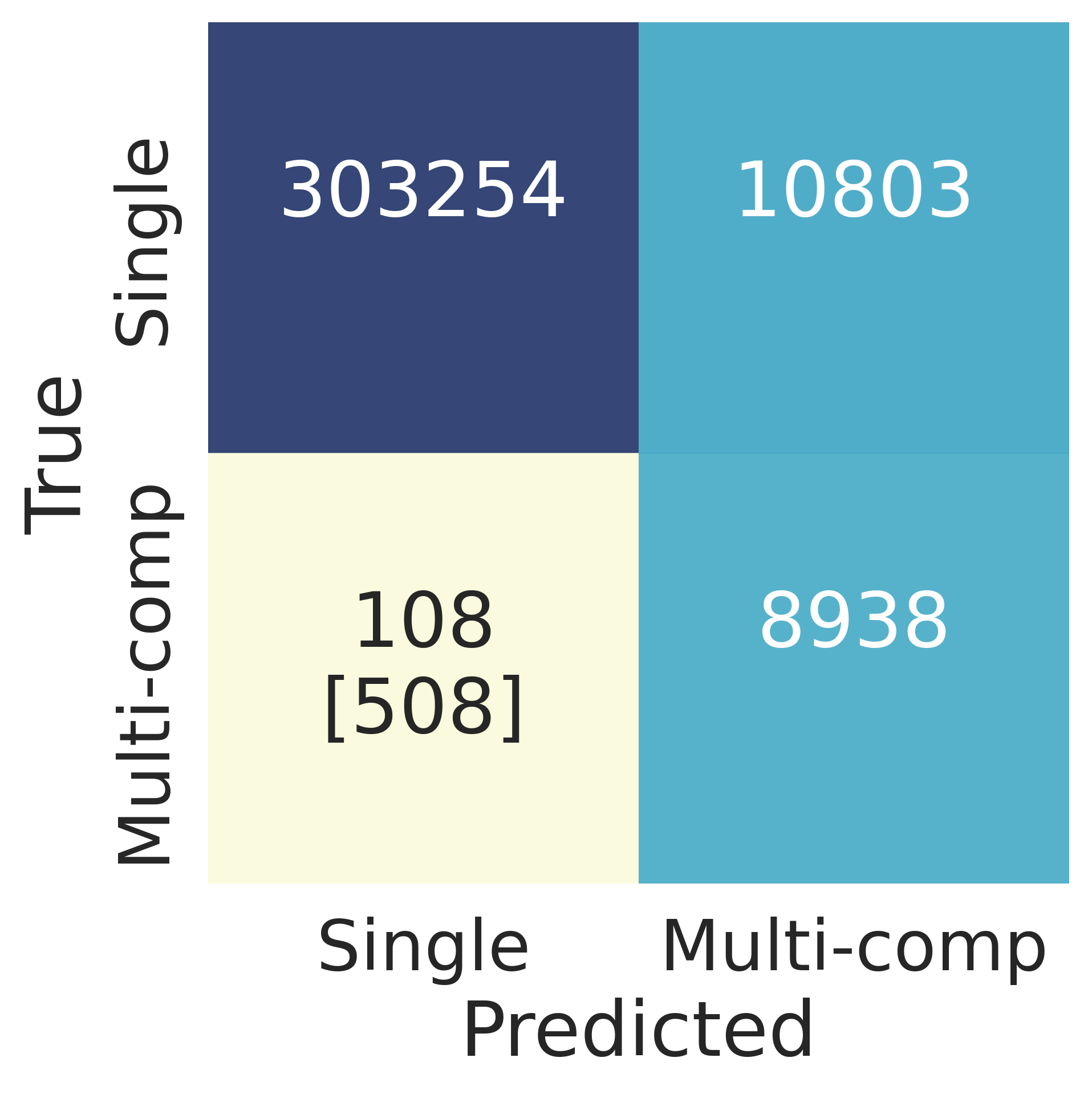}
&
\includegraphics[scale=0.33]{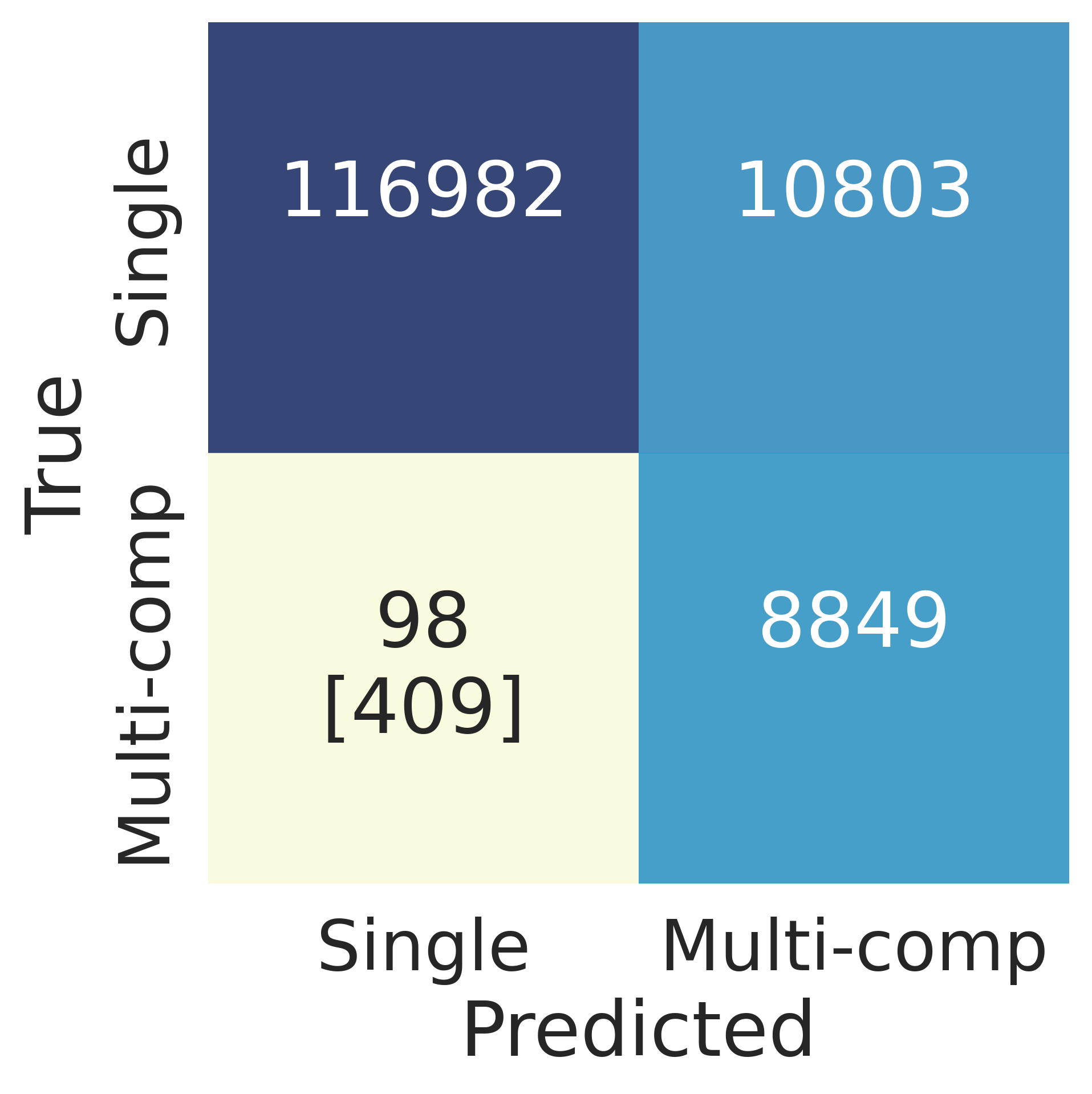}
&
\includegraphics[scale=0.33]{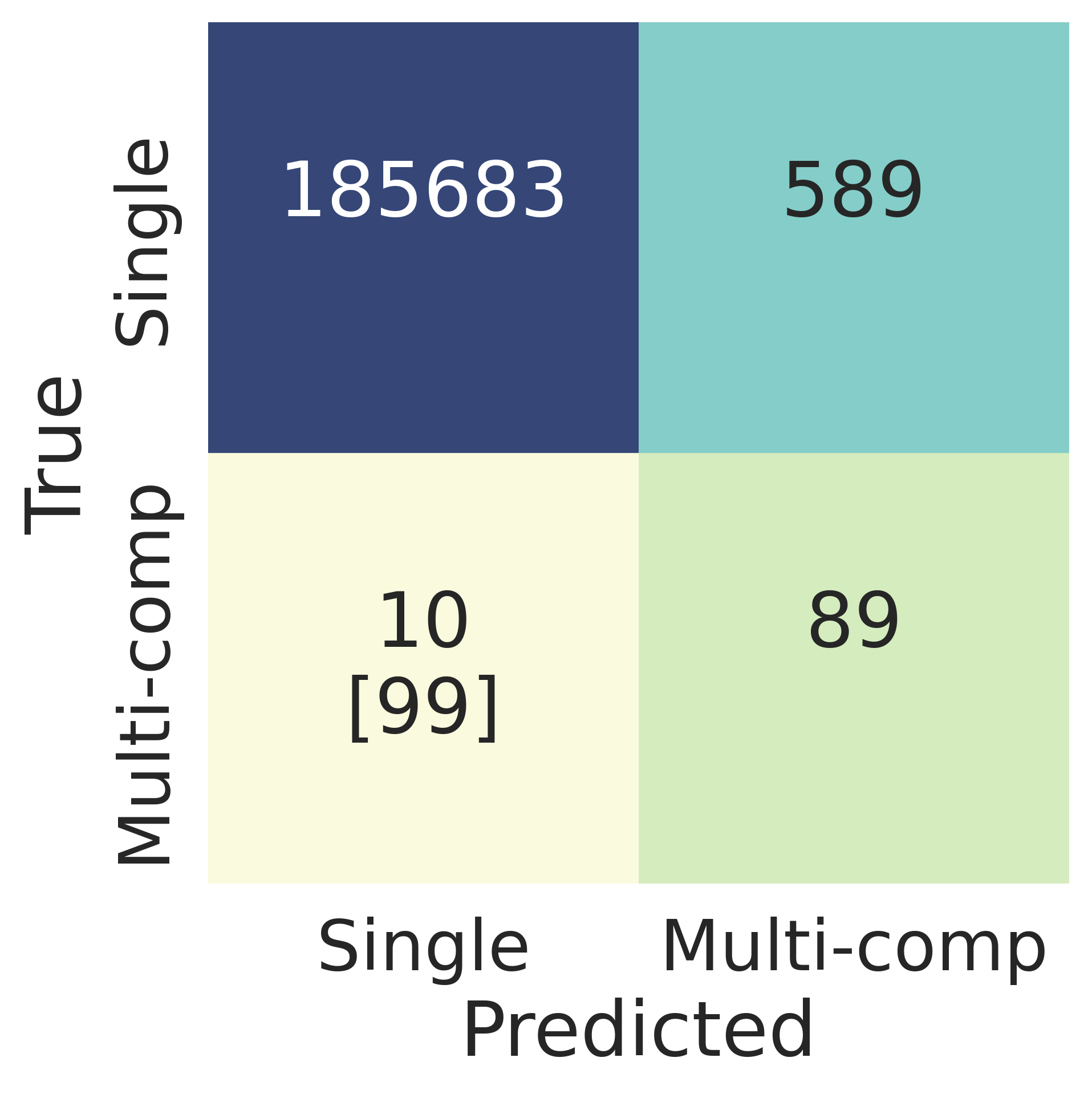}
\end{tabular}
\end{minipage}
\caption[Confusion matrix for LoTSS DR1]{Confusion matrix for all the sources in LoTSS DR1 using the final adopted model and a threshold value of 50 per cent. Left: results for all the sources in LoTSS DR1 setting the SISG to class S. Middle: results on DR1 with the SISG sources removed. Right: results for the SISG sources only. In all panels, the values in the square brackets correspond to the numbers of FP before applying the corrections. }
\label{Fig:CM_full_ds_individuals}
\end{figure*}

Using the adopted threshold value of 0.5~\footnote{In the online table, the column \textit{mc$\_$}prediction$\_$0.5 corresponds to the predictions for a threshold value of 0.5, with the following values: (0) sources predicted as class S; (1) sources predicted as class MC; and (2) sources corrected (i.e. recovered to class MC) as described in Section~\ref{sec:performance_threshold}. The actual prediction values correspond to the \textit{mc$\_$}probability$\_$multi column.}, we analyse the performance of the classifier across the entire dataset and on different categories of sources. This is done by analysing the results of the Confusion Matrix (CM; see Appendix~\ref{sec:ML_metrics}) where the values on the CM correspond to the number of sources in the TP, TN, FP, and FN classes, defined earlier in this section.
The results of the model applied to the full LoTSS DR1 sample can be seen in Figure~\ref{Fig:CM_full_ds_individuals}. The figure also compares how the model performs when confronted with the SISG, which were excluded during training. As will be explained next, the adopted strategy will consist of training the model with the SISG removed and applying it to all DR1 sources except for SISG, and then setting the SISG to class S (i.e. sources that are automatically classified as not being MC sources).

The left CM in Figure~\ref{Fig:CM_full_ds_individuals} shows the results of the final model (in which all SISG sources are assigned to class S, i.e. not MC sources), while the middle and right CM correspond to the results when the final model is applied to non-SISG and SISG sources, respectively. These result from training the model with SISG sources removed, calculating a prediction for all LoTSS DR1 sources, and separating the data by SISG sources. The left CM is subtly different from the sum of the values in the different cells of the middle and right CMs because SISG sources were all automatically set to class S. Therefore, all the SISG sources that had been classified correctly (89 sources) or incorrectly (589 sources) as MC sources will contribute to the values in the left column in the left-hand CM. By setting SISG to class S, the classification is improved by saving almost 600 sources from the FP, even though 10 more sources (after correction) end up as a FN. However, this represents a good trade since it means a maximum of 5 physical radio sources because, by definition, each MC is made up of at least 2 source components.
Using the adopted strategy and the 0.5 threshold, the accuracy of the model when applied to the imbalanced LoTSS DR1 dataset is 96.62 per cent.

This demonstrates that the overall results when applying the model to other data are also improved if the SISG sources are set automatically assigned to class S. Based on this conclusion, the SISG sources can also be excluded from the data processing (and its predictions set to class S). This results in only about 40 per cent of the data requiring to be processed.

\subsection{Performance as a function of sources properties}

\begin{figure*}
\begin{minipage}{\textwidth}
\begin{tabular}{lll}
\centering
\includegraphics[scale = 0.31]{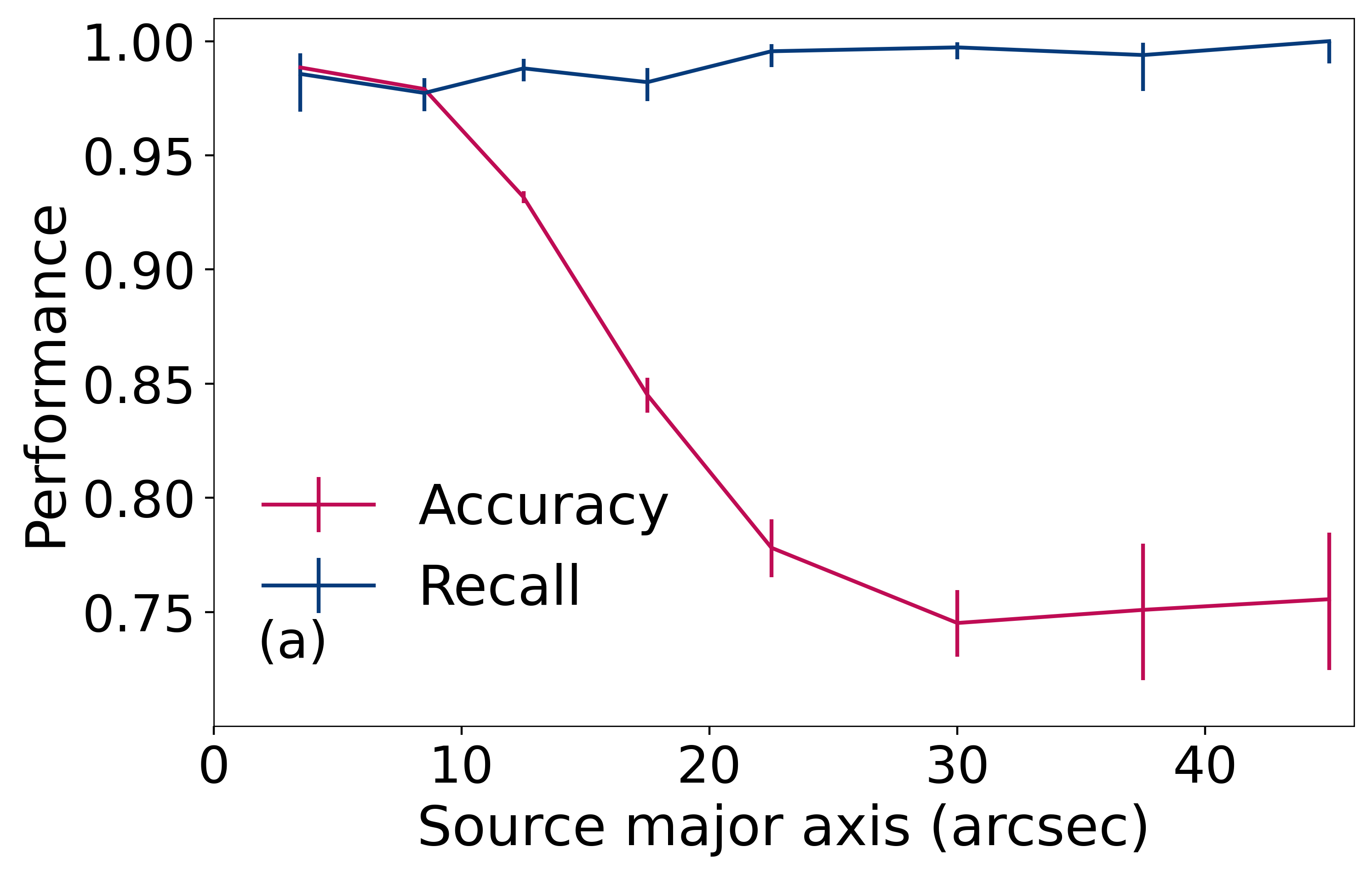}
& \includegraphics[scale = 0.31]{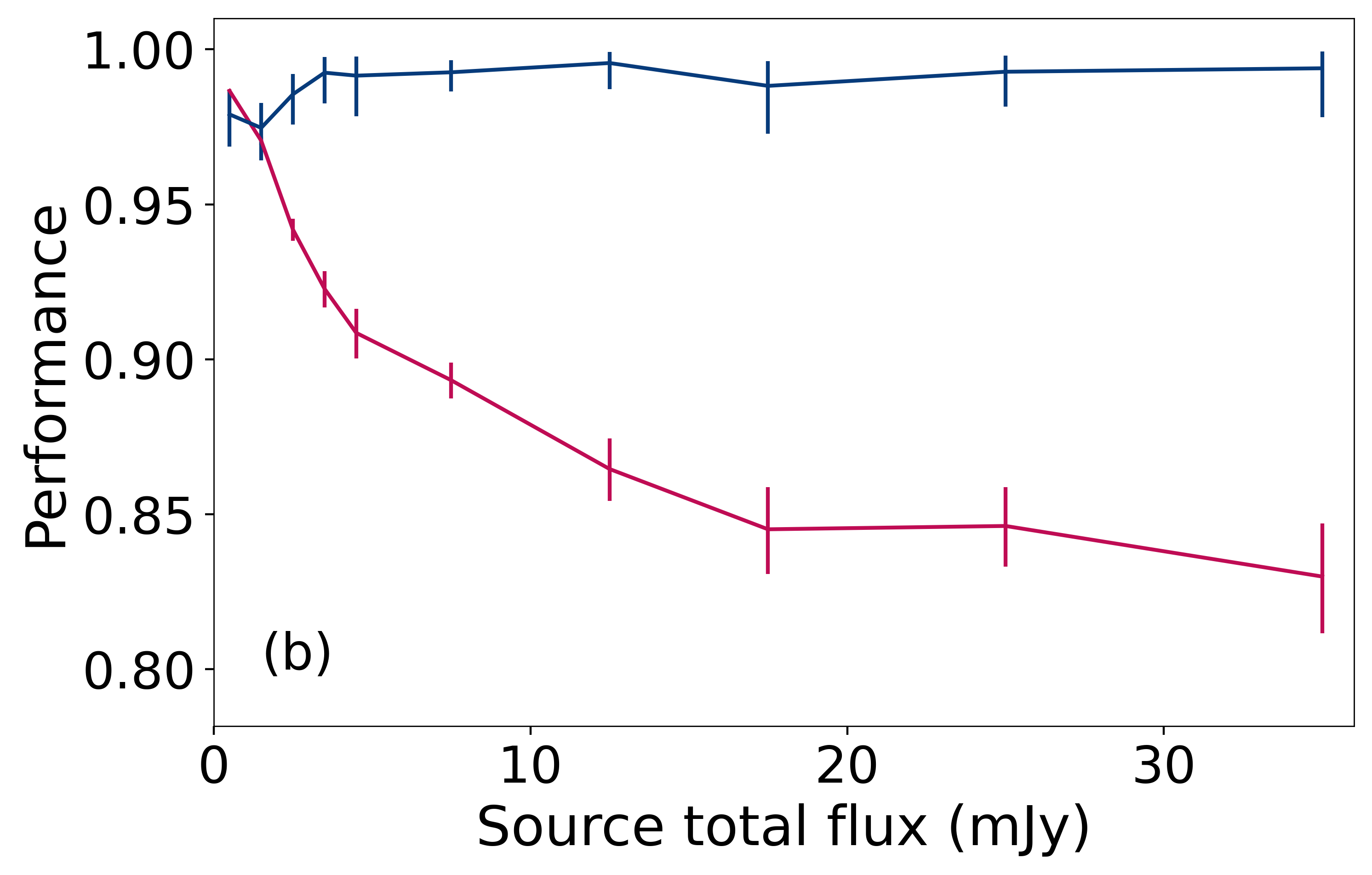}
& \includegraphics[scale = 0.31]{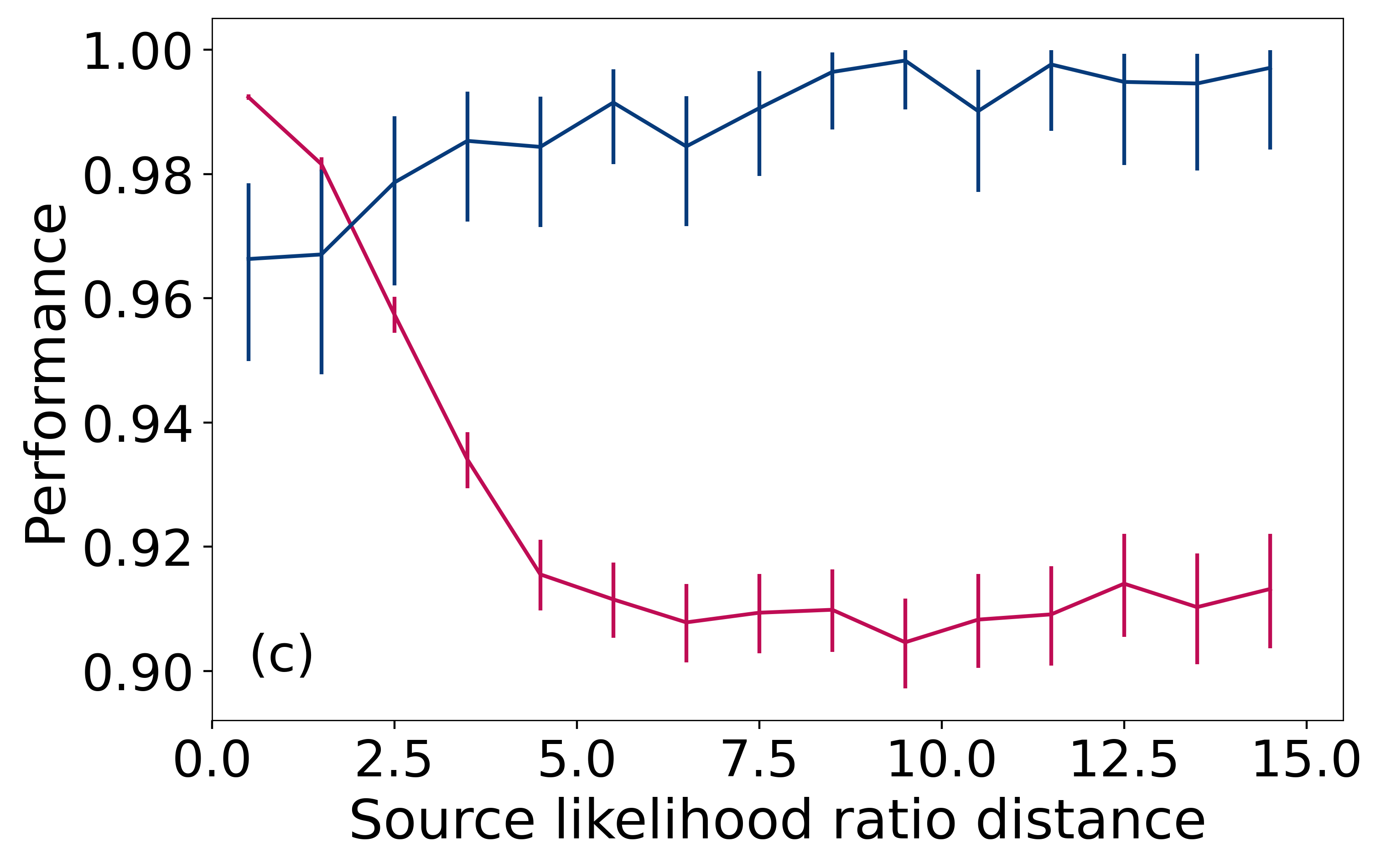}\\
\includegraphics[scale = 0.31]{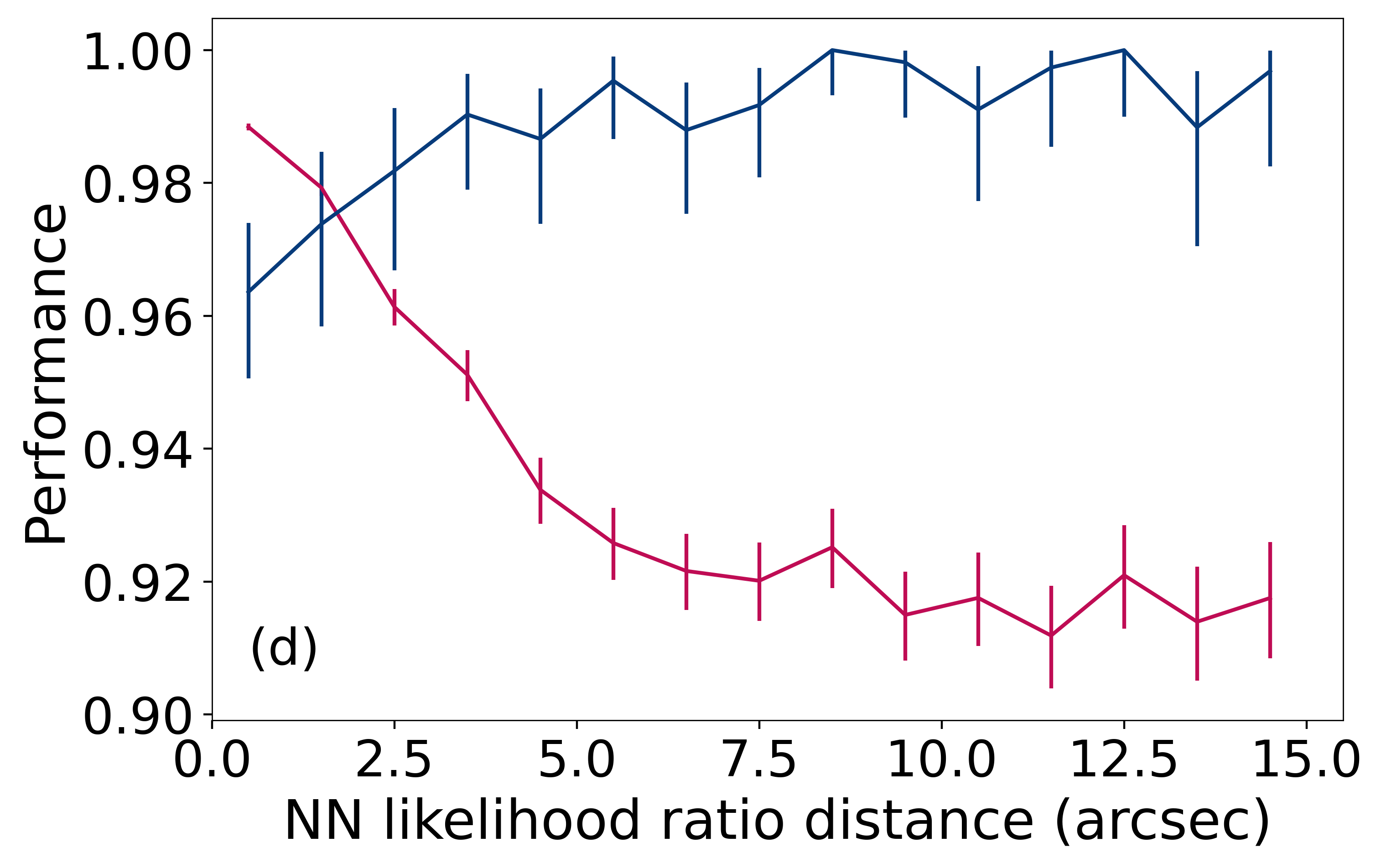}
& \includegraphics[scale = 0.31]{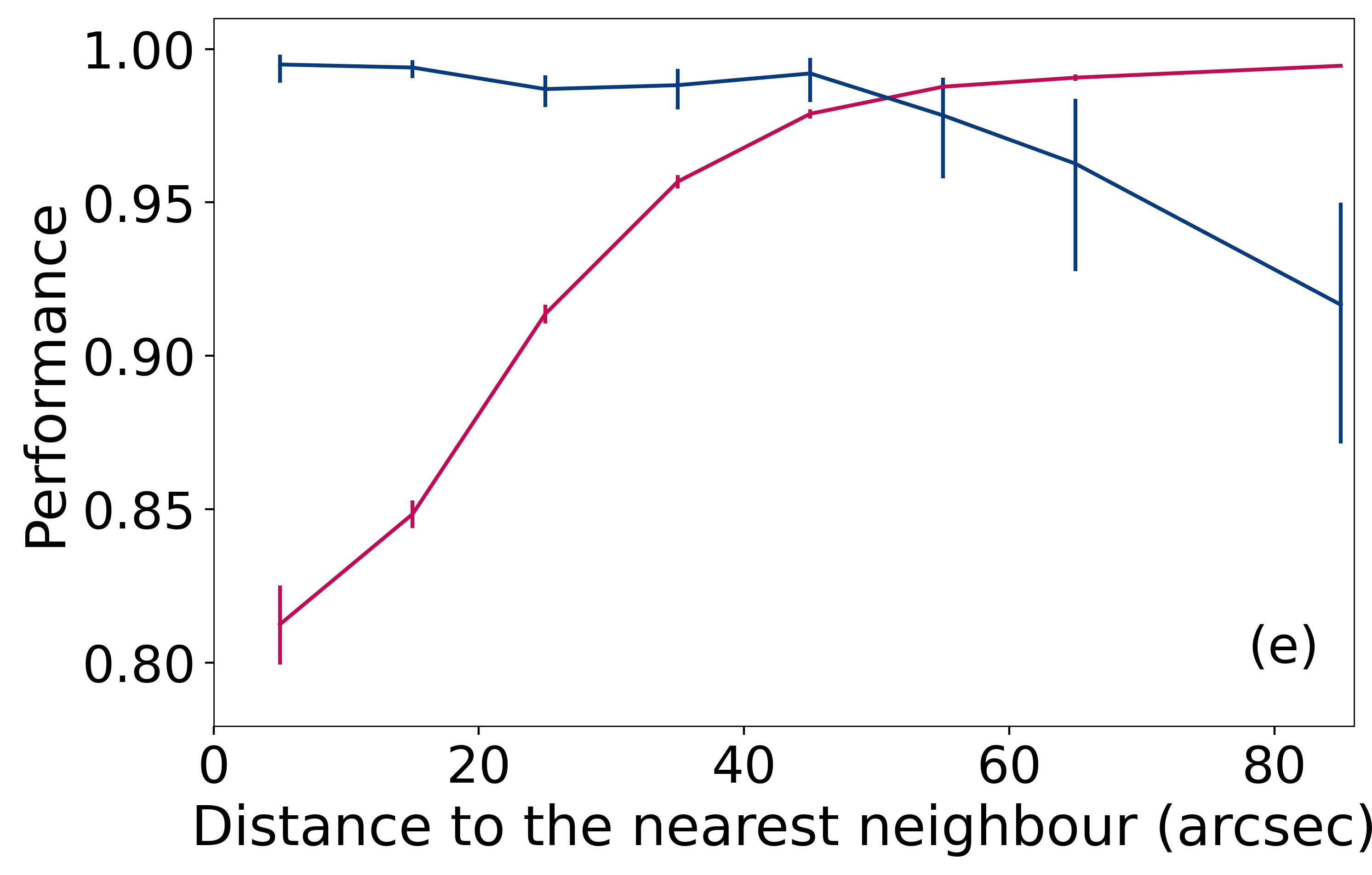}
& \includegraphics[scale = 0.31]{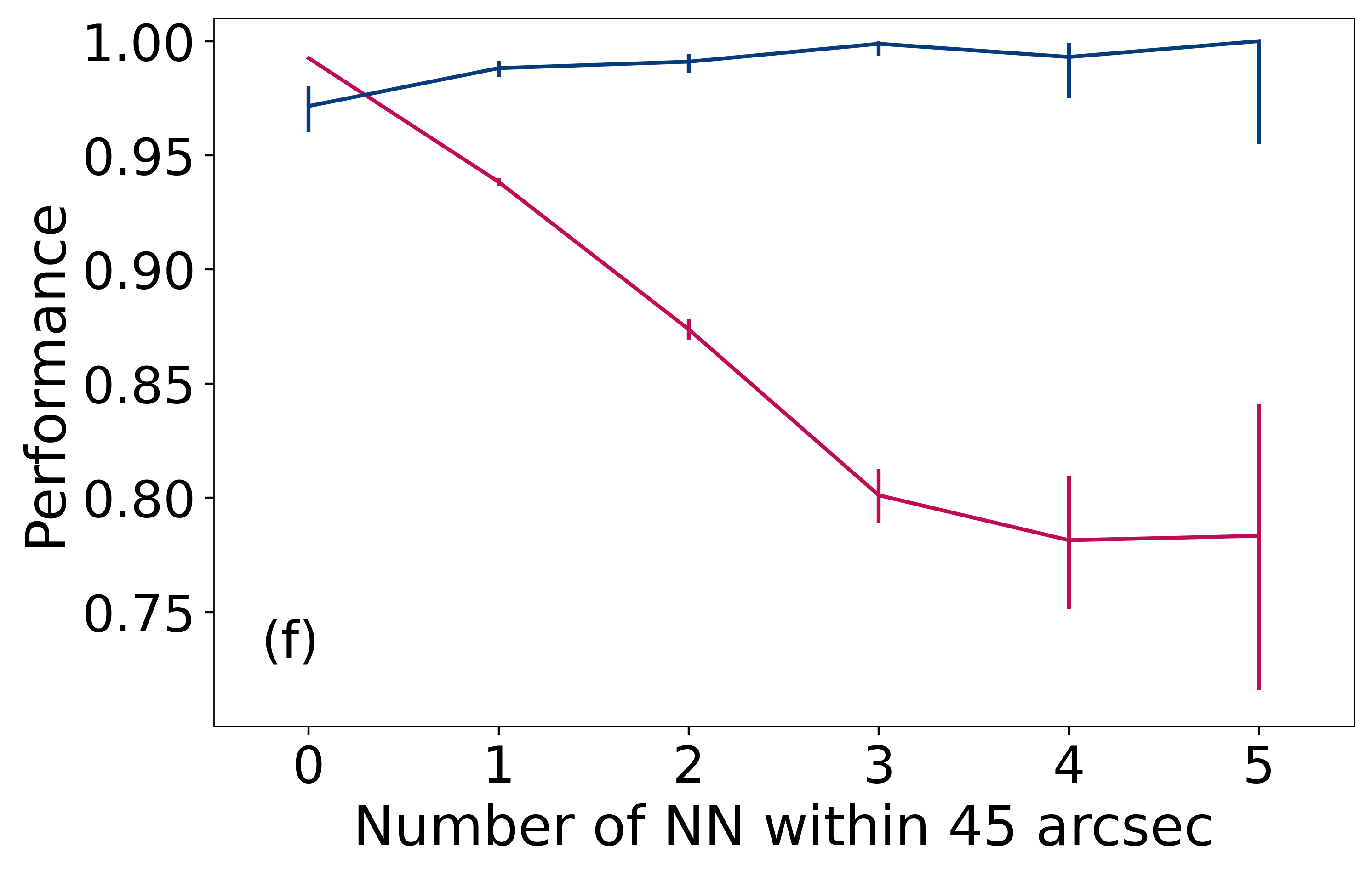}
\end{tabular}
\end{minipage}
\caption[Performance as a function of radio source properties]{Performance as a function of radio source properties, with accuracy displayed in red and recall for the MC class in blue. The histograms show that the final model adopted has high accuracy across the different properties being analysed, particularly for smaller and fainter sources and those with no near neighbours. The values of recall, however, are always significant above 0.95, with the sole exception of when there is no nearest neighbour within 60 arcsec. High values of recall are due to a consistently high number of sources being identified as MC sources and a low number being missed. The values of precision (not plotted) are consistently weaker and have values around 0.5 across all the parameter space, indicating a large number of false positive (as also seen in the confusion matrix). See the text for a discussion.}
\label{fig:hist_results}
\end{figure*}

In order to understand the performance of the model and its ability to distinguish between class MC and class S, we evaluate the performance of the classifier as a function of source characteristics and contextual information. This is illustrated in Figure~\ref{fig:hist_results}. 

Regarding the angular sizes of the \texttt{PyBDSF} source being analysed (panel a), accuracy is consistently above 95 per cent for sources with major axis up to around 10 arcsec, indicating a successful identification of class S sources through a high number of TN. These type of sources correspond to the majority of sources in the LoTSS surveys. Interestingly, even at these small angular sizes where single-component sources dominate the sample, the recall for MC sources remains high. 
The accuracy drops steeply as the source size increases, reaching around 75 per cent for sources with 25-30 arcsec and remaining at this value for larger sources (albeit that there are relatively fewer sources of this size in the sample). The drop in accuracy for larger sources is primarily due a decrease in the proportion of TN, that is, sources that are actually single component being correctly identified, because there are less single component sources with these sizes. 

The accuracy is above 90 per cent for sources with total flux density (panel b) below 4-5 mJy, but it drops for brighter sources, particularly for sources brighter than 15 mJy, where the performance drops to about 85 per cent. High performance at lower flux densities is due to a high number of TN since the majority of the sources in LoTSS are faint. 
Interestingly, a higher proportion of FP can also be found in the fainter bins, with extended and faint emission being more likely to be part of a MC source as opposed to bright emission.
Lower performance at higher flux densities is attributed to a small fraction of TN. Recall of MC sources remains consistently high at all flux densities.

The classifier shows high performance values for sources where the distance to the LR counterpart (panel c) is low, in particular for values below 1-2 arcsec. Small values of LR match distance (and high values of LR value) indicate a genuinely-associated optical match, suggesting that the source is a single-component source in most cases (the alternative being the core of a MC source). Therefore, the classifier is able to correctly identify these sources as class S sources. The accuracy drops sharply for higher values (but it is always above 90 per cent) as the proportion of TN falls. The highest proportion of sources being misidentified as MC sources can be found for the 3 smaller LR distances bins. A similar conclusion can be drawn if inspecting the performance using the nearest neighbour (NN) LR distance (panel d). Smaller NN LR distances have a higher probability of the NN source being a MC source, and therefore a high probability of the source being a MC itself, since it takes two source components to make up a MC source. The performance drops for higher values of the NN LR distance, as happens with the source LR distance, due to a drop in the TN. In all cases the accuracy is always above 90 per cent for higher LR distance matches.

The performance as a function of the NN properties (panel e) is evaluated further since the presence of a NN is an indication that the source might be clustered and potentially has a higher chance of being part of a MC source. If the first NN is more than around 50 arcseconds apart, the accuracy is close to 100 per cent, indicating that the majority of these sources do not need to be grouped and are correctly identified by the model as class S sources. The recall of MC sources for such distance nearest neighbours is at its lowest here of all of the parameter space examined in Figure~\ref{fig:hist_results}, but still remains above 90 per cent. 
Smaller distances to the NN suggest a more crowded environment and increase the chances of the source being a MC source, and therefore the accuracy of the classifier drops due to the mixed population. 

When evaluating the performance as a function of the number of NNs within 45 arcseconds (panel f), it is possible to observe that the classifier reaches accuracy values close to 100 per cent when there are no NNs within this radius because the chances of being a class MC source are comparatively low, and there is an accurate identification of the class S sources, for which the majority do not have any NN within 45 arcseconds. As the number of NNs within the radius increases, the accuracy drops, mainly because there are a smaller number of single-component sources in these bins.

\subsection{Performance below 4 mJy flux density}
\label{sec:4mjy}

\begin{figure}
\centering
\begin{tabular}{ll}
\hspace{1.2cm} Sources <~4 mJy & \hspace{1.2cm} Sources >~4 mJy \\
\includegraphics[width=0.45\columnwidth]{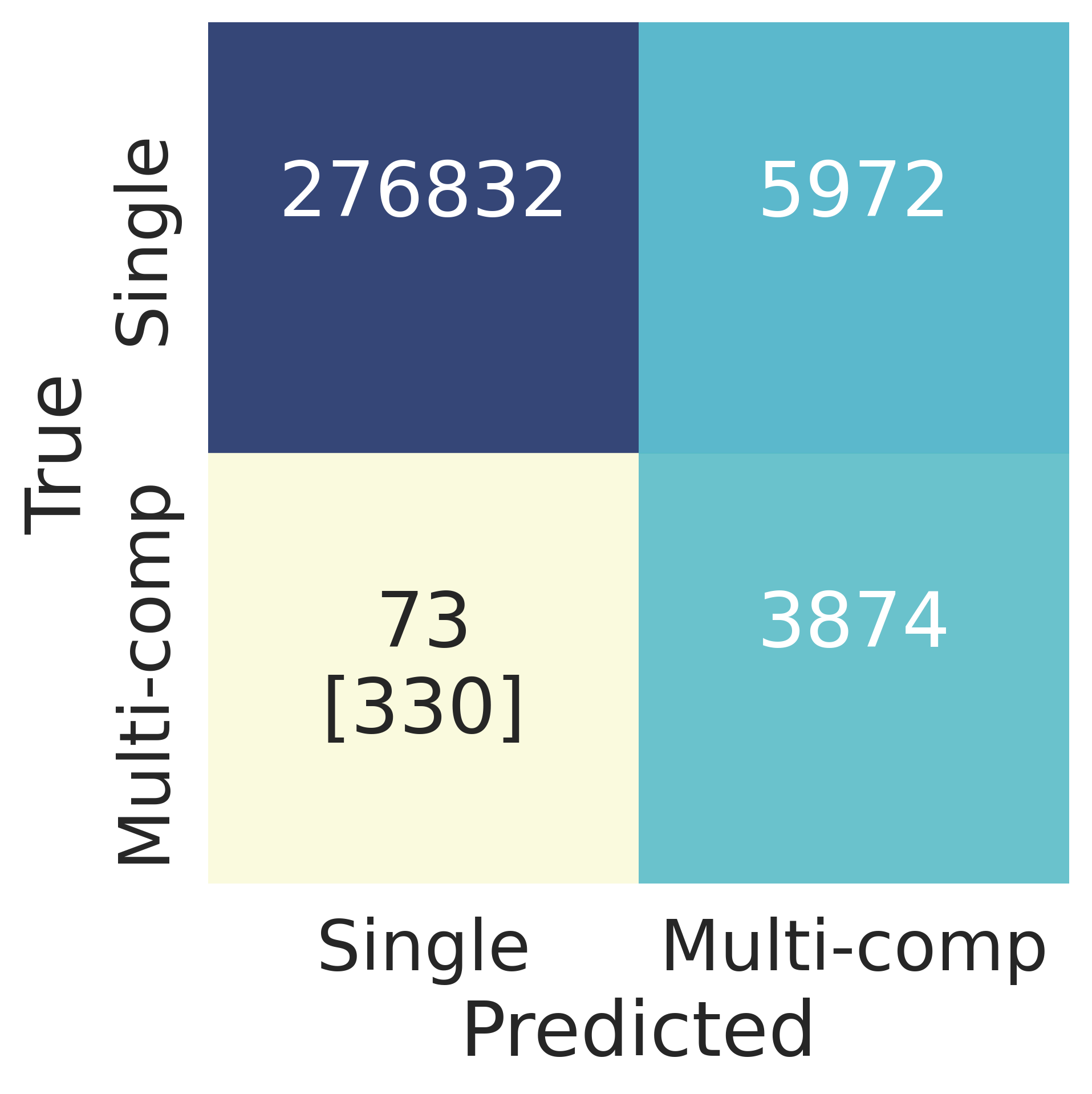}
&
\includegraphics[width=0.45\columnwidth]{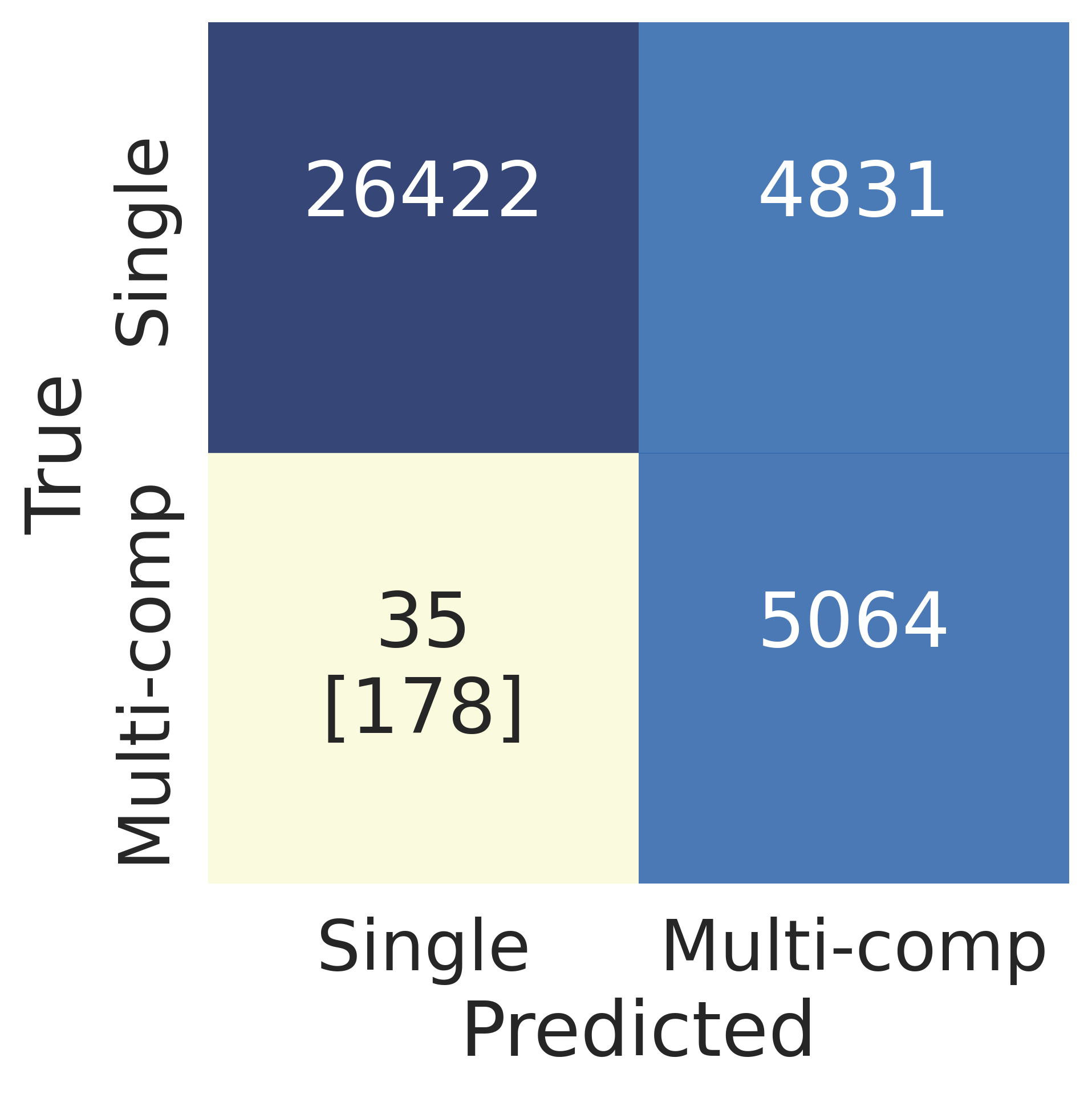}
\end{tabular}
\caption[Confusion matrices for 4 mJy sources]{Confusion matrices for the entire LoTSS DR1 dataset, for sources below (left) and above (right) 4 mJy. }
\label{Fig:cm_4mjy}
\end{figure}

Due to the amount of sources in LoTSS, in LoTSS DR2 sources with a flux density of less than 4 mJy were not sent for visual inspection \citep{Hardcastle2023dr2}. This is mainly because priority was given to potential WEAVE-LOFAR \citep{smith2016weave, Jin2023}  target sources (which are brighter than 8 mJy) for spectroscopic follow-up. Furthermore, there are many sources below 4 mJy, most of which are single-component sources \citep[see][]{williams2019lofar}.
Those faint sources which are multi-component are, in general, very difficult to identify, which would represent a huge effort without much return. Hence they have not been inspected in DR2, and therefore, it is important to investigate the performance of the model below this 4 mJy threshold. The ability for the classifier to identify MC sources for fainter (below 4 mJy) and brighter (above 4 mJy) sources can be seen in Figure~\ref{Fig:cm_4mjy}.

The classifier demonstrates higher accuracy when classifying fainter sources, achieving 97.89 per cent accuracy. The accuracy drops to 86.61 per cent for brighter sources. There is a much larger population of sources below 4 mJy compared to those above it, and the large values of accuracy for faint sources are because a considerable number of them are correctly classified as single-component sources. The number of sources flagged as MC sources (both correct and incorrect classifications) has a similar order of magnitude for sources above and below 4 mJy. Therefore, the performance of the classifier is comparable among these two groups, with the large majority of the genuine MC sources being correctly flagged as MC, and a similar number of sources being incorrectly flagged as MC. This is more pronounced for fainter sources, but without major differences. Furthermore, the model successfully identifies nearly all the sources that necessitate component association, missing less than 2 per cent of those even if the source is faint.

\begin{figure*}
\begin{minipage}{\textwidth}
   \centering
\begin{tabular}{cc}
\includegraphics[width=0.4\columnwidth]{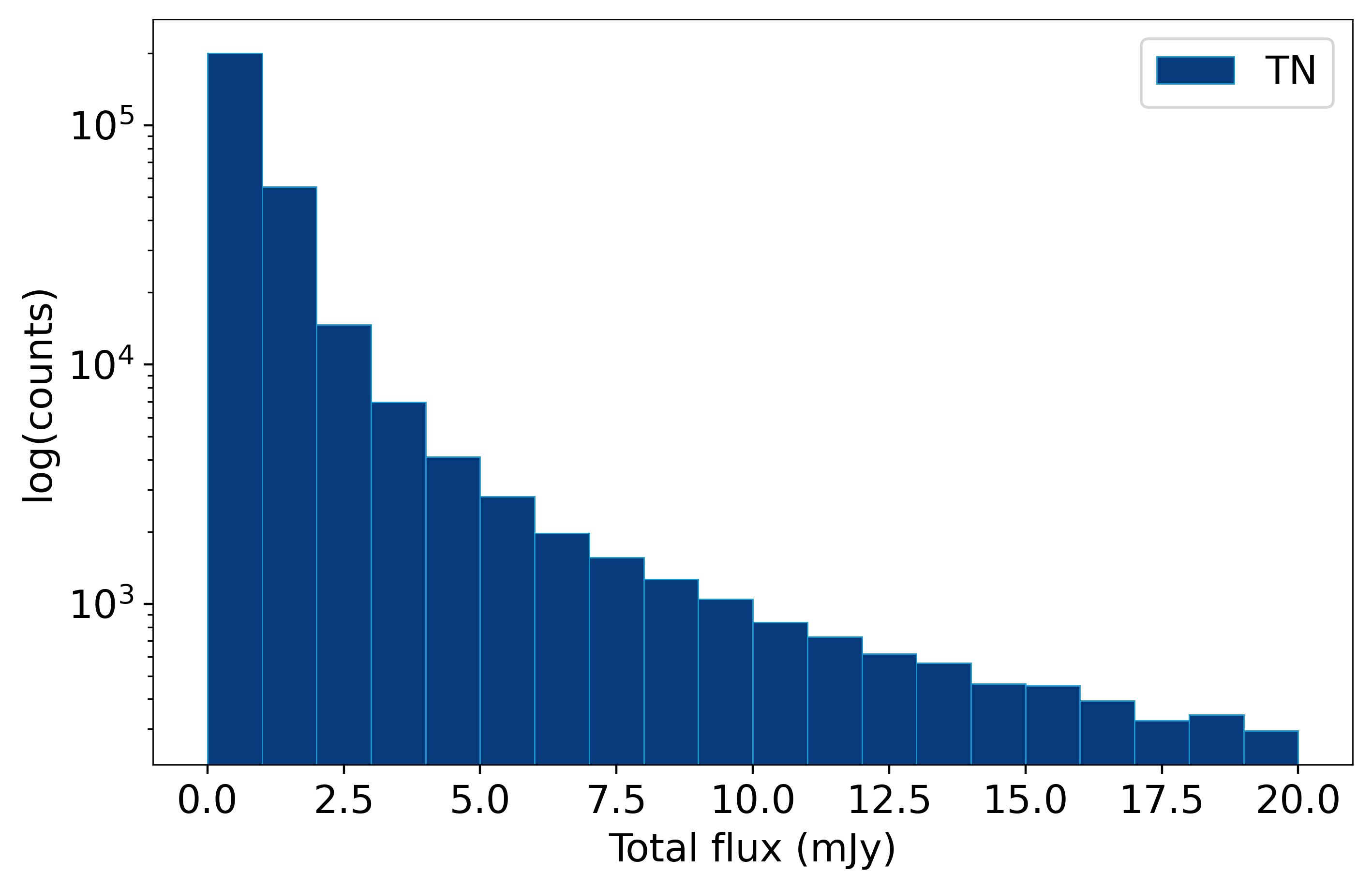}&
\includegraphics[width=0.4\columnwidth]{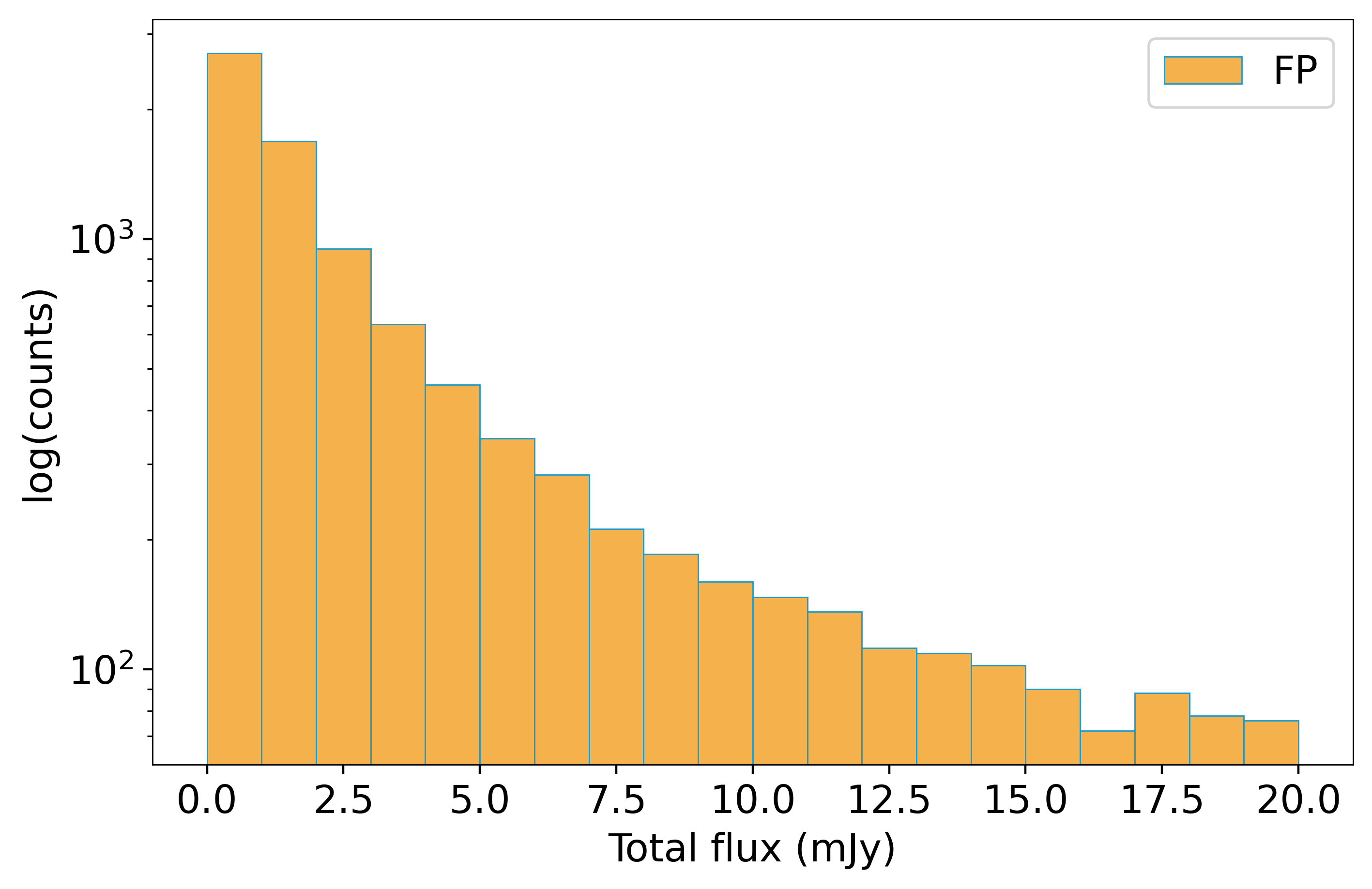}\\
\includegraphics[width=0.4\columnwidth]{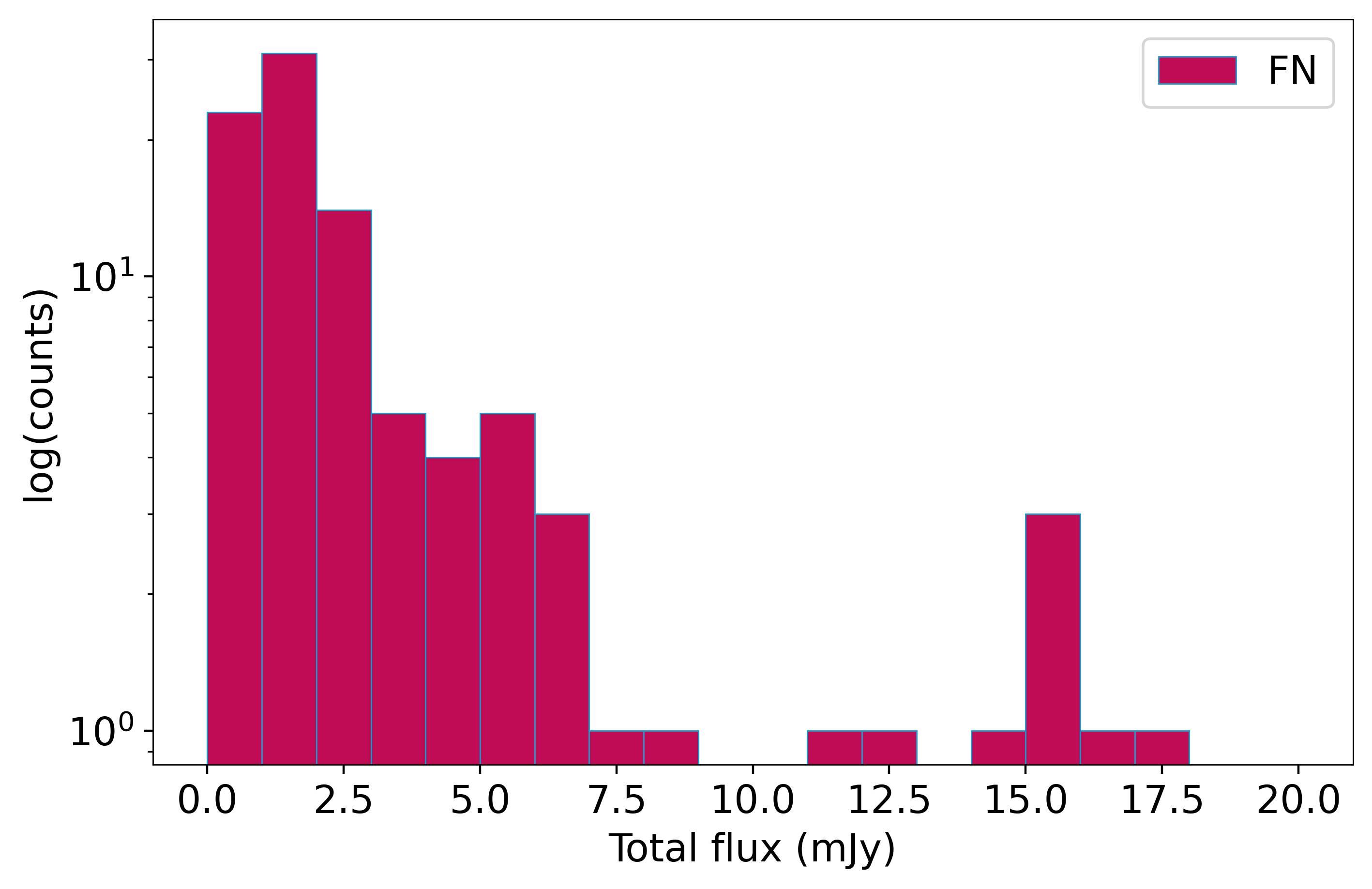}&
\includegraphics[width=0.4\columnwidth]{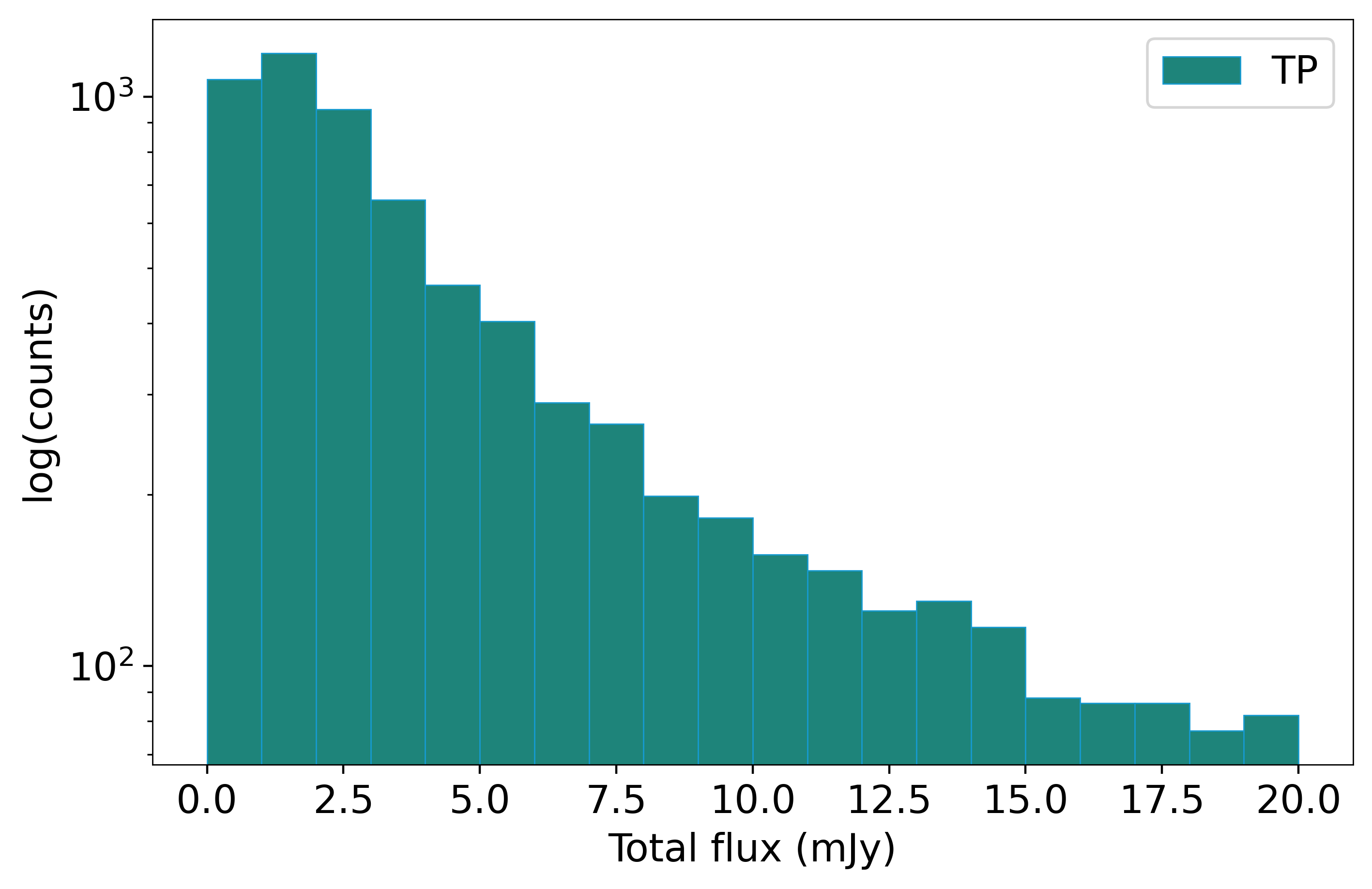}\\
\end{tabular}
\end{minipage}
\caption[Confusion matrix histograms in functions of total flux density]{Confusion matrix counts in terms of flux density. Each of the cells correspond to: TN (top left, blue), FP (top right, yellow), FN (bottom left, red), and TP (bottom right, green).}
\label{fig:flux_cm} 
\end{figure*}

The distribution of sources in each of the cells of the confusion matrix as function of the flux density can be seen in Figure~\ref{fig:flux_cm} (note the logarithmic y-axes). At lower flux densities, the abundance of single-component sources is higher, and the number of correctly classified class S sources (TN) is also higher. There is a reduction in the number of TN sources as the flux density increases, but this is because there are fewer bright sources overall.

For sources with lower flux densities, there is a greater number of FN, but at high flux densities a higher proportion of sources are multi-component compared to at low flux densities. This is also why accuracy drops at high flux densities (see Figure~\ref{fig:hist_results}).
This trend is also evident when examining the distribution of sources in the TP and FP histograms. The occurrence of FP is predominantly observed at lower flux densities, but this is also because there are many more sources at these flux values.

\section{Conclusions and future outlook}
\label{sec:conclusions}

The number of faint sources with intricate radio structures is increasing in modern radio continuum surveys. Sometimes source components can be mistakenly identified as independent sources despite being components of the same physically connected radio source. This work introduces a multi-modal deep learning classifier specifically designed to identify these MC sources. These are sources that require component association and for which currently there are no automatic identification methods available.
This work has implications for future surveys as it becomes impractical to select and cross-identify all sources using conventional astronomy techniques, which commonly involve substantial amounts of visual analysis. The work also highlights the effectiveness of deep learning algorithms, particularly when combining data from diverse sources, as a valuable approach for handling modern radio surveys. 

The model developed in this work combines a convolutional neural network and an artificial neural network into a single architecture. The model incorporates radio images and source parameters of the radio sources and their nearest neighbours, as well as parameters of the possible optical statistical counterpart. The model is trained using LoTSS DR1 manual annotations to discriminate between a) sources that are part of MC sources (which will always be difficult to identify and cross-match) and b) relatively compact sources, which can be processed in a more automatic way using statistic methods or machine learning methods such as those of \citet{Alger2018}, and are also typically unresolved single-component sources.
We used 9,046 MC \texttt{PyBDSF} sources out of the total 323,103 \texttt{PyBDSF} sources identified in LoTSS DR1. While 75 per cent of the MC sources were used for training purposes, 30 per cent was split equally for validating and testing the model. The dataset was augmented by performing rotations and flips on the MC sources and using a proportional number of random single-component sources, in order to achieve a balanced dataset. The dataset after augmentation comprised 42,946 sources, of which 37,578 were used for training, 2,685 for validation and 2,683 testing (we defined the validation and test sets to be the same size as the ones before augmenting the training set).
In this work, we employ active learning by excluding SIGS sources from the dataset before the training process. These sources do not add diversity to the dataset and can be predominantly cross-matched using statistical methods. By removing the SIGS sources, we increase the ability of the model to detect MC sources and save processing time since these correspond to approximately 60 per cent of the LoTSS data. 

The model demonstrates good results, achieving a recovery rate of 94 per cent for sources with MC in the balanced dataset and an overall accuracy of almost 97 per cent in the real imbalanced dataset consisting of 323,103 sources. The performance of the classifier is closer to 100 per cent for small, and faint sources, dropping for sources brighter than 2-3 mJy and sources larger than 10 arcseconds. The classifier shows excellent performance (between 96 per cent and 99 per cent) for sources with smaller distance to an optical counterpart, in particular if the source itself or the NN have a LR match below 1-2 arcseconds, which is an indication that the source (and its NN) are not part of a MC source. The classifier precisely identifies class S sources with 99 per cent accuracy if there are no NN within 45 arcseconds. Furthermore, if the NN is smaller than 10 arcseconds the classifier performs closer to 98 per cent. We evaluated the performance of the classifier for sources below 4 mJy since those are not being visually inspected in LoTSS DR2 \citep{Hardcastle2023dr2}, and a good performance is achieved for both brighter (86.6 per cent accuracy) and fainter (97.9 per cent accuracy) regimes, with many more fainter sources being correctly identified as class S sources since the majority of the sources in LoTSS are indeed faint and single-component sources. These results indicate that the reliability of the classifications heavily depends on the distribution of the source characteristics within the dataset. 

Our model already exhibits strong performance. However, deep learning is a flourishing field, with new architectures and methods being developed rapidly, and there are a variety of ways in which the model could potentially be improved. Investigating different types of fusion could lead to improvements, for example the architecture could implement a fusion module where the weights of the CNN and ANN are shared across the network instead of performing a single late fusion. Another option could be to construct an ensemble of classifiers to enhance the model's performance, which could be done with any other type of machine learning or deep learning model. Furthermore, the architecture could be optimised using AutoML, which would help automate the network design process and optimise hyperparameters. Conducting feature exploration, such as grouping features or designing new features, could improve the ANN part of the model. Finally, incorporating different wavelength images could be explored, such as optical and infrared, although their impact is expected to be more important for source cross-matching than for this source classification task in particular.
The construction of the dataset could be evaluated in order to examine the performance of the model when blended sources are in the same class as MC sources or when there is discrimination between those three independent classes. This would raise the question of whether the radio source detector was accurate in identifying the source itself. Furthermore, it would be interesting to assess whether additional training examples improve the overall performance, which could be achieved using the outputs from the citizen science annotation of LoTSS DR2 \citep{Hardcastle2023dr2} to train and evaluate the model.

\citet{Mostert2024} assembled a pipeline to automatically group and cross-match multi-component radio sources. The source association part of the pipeline builds on the approach of \citet{Mostert2022} for component association. However, while that algorithm performs well on genuine MC sources, if single-component sources are included, then 7.7 per cent of them get erroneously grouped with unrelated \texttt{PyBDSF} sources. 
In addition, they assume that the majority of their galaxies will belong to the type of sources identified by \cite{alegre2022} as the ones that cannot be matched using the LR technique. While this is expected, \cite{alegre2022} do not specifically address whether a source requires radio component association. The present work will help to tackle this question by determining the specific subset of sources on which the source association code should be executed. This will also allow the pipeline to be expanded to include fainter and smaller sources than it does now.
Our results will therefore improve the overall pipeline for automatic source association and identification in LoTSS.
The proposed methodology would involve three main steps. 
Firstly, the findings of the present study are used to identify the \texttt{PyBDSF} sources that are most likely to be part of a MC source. Secondly, the \citet{Mostert2022} component association code is executed to define the physical radio sources (possibly extending the method to smaller and fainter sources).
This uses the output of \cite{alegre2022} to eliminate unrelated single-component sources within the bounding box of the extended source, for which the threshold value can be adjusted as well. Finally, after the sources have been associated, the \citet{barkus2022application} code is used to obtain the optical identifications using the ridgeline approach.

In conclusion, in LoTSS DR1 and LoTSS DR2, a substantial effort was put into analysing the sources that require component association. This was done manually on LGZ by associating components and cross-matching. Therefore, the outcomes of this work are of significant value for incorporating into pipelines for the processing of upcoming LoTSS data releases or other radio surveys. Furthermore, the results can be incorporated into diverse pipelines not only for automated cross-matching but also for identifying sources for further radio morphology classification or for the simple detection of radio sources (for example, by ensuring the radio properties correspond to actual sources).


\section*{Acknowledgements}

LA is grateful for support from the UK Science and Technology Facilities Council (STFC) via CDT studentship grant ST/P006809/1.
PNB and JS are grateful for support from the UK STFC via grants ST/R000972/1 and ST/V000594/1. The authors would like to express their gratitude to the referee for the valuable feedback that improved the clarity of the paper. The authors thank Deyan Petrov and Adam McCabe for their help in the early stages of this project.
LOFAR data products were provided by the LOFAR Surveys Key Science project (LSKSP; \hyperlink{https://lofar-surveys.org}{https://lofar-surveys.org}) and were derived from observations with the International LOFAR Telescope (ILT). LOFAR \citep{vanHaarlen2013lofar} is the Low Frequency Array designed and constructed by ASTRON. It has observing, data processing, and data storage facilities in several countries, that are owned by various parties (each with their own funding sources), and that are collectively operated by the ILT foundation under a joint scientific policy. The ILT resources have benefitted from the following recent major funding sources: CNRS-INSU, Observatoire de Paris and Universit{\'e} d'Orl{\'e}ans, France; BMBF, MIWF-NRW, MPG, Germany; Science Foundation Ireland (SFI), Department of Business, Enterprise and Innovation (DBEI), Ireland; NWO, The Netherlands; The Science and Technology Facilities Council, UK; Ministry of Science and Higher Education, Poland.


\section*{Data availability}

The datasets were derived from LoTSS Data Release 1 publicly available at https://lofar-surveys.org/dr1$\_$release.html. The table of source features derived for this work is provided as supplementary online material, along with the model predictions. The model is available by request to the lead author.


\vspace{-0.5cm}
\renewcommand*{\bibfont}{\footnotesize}
\setlength{\bibsep}{1pt}
\bibliographystyle{mnras} 
\bibliography{bibliography.bib}



\appendix
\section{Performance Metrics for supervised classification}
\label{sec:ML_metrics}

In classification problems, each example belongs to one of several classes. Binary classification has two classes, which are commonly labelled as positive and negative (or 1 and 0). Table~\ref{tab:ml_cm} presents the ``confusion matrix'' for a binary classification problem, where the true positive \textit{TP} and the true negative \textit{TN} are the number of values which are correctly identified by the classifier, from the positive and negative classes, respectively. The false positive \textit{FP} and false negative \textit{FN} correspond to the remaining number of values classified as positive and negative, respectively, but which belong to the opposite class. The confusion matrix may be used to derive standard metrics by which the performance can be evaluated \citep[see e.g.][for a review]{hossin2015review}. 

\begin{table}[h]
\centering
\begin{tabular}{l|l|c|c|c}
\cline{3-4}
\multicolumn{2}{c|}{}&Positive&Negative\\
\cline{2-4}
\multirow{1}{*}{\rotatebox{90}{True}} 
& Positive & TP & FN \\
\cline{2-4}
& Negative & FP & TN \\
\cline{2-4}
\multicolumn{2}{c}{}&\multicolumn{2}{c}{Predicted}&\\
\end{tabular}
\caption{Binary classification confusion matrix.}
\label{tab:ml_cm}
\end{table}

Accuracy is the most popular performance metric. It measures the fraction of sources which are correctly classified relative to the overall classifications: 

\begin{equation}
  Accuracy = \frac{TP+TN}{TP+TN+FP+FN}  
\end{equation}

\noindent When using a balanced dataset (i.e. when each of the classes has a similar number of examples), accuracy shows how well the classifier performs overall. However, for imbalanced datasets, the accuracy may not reflect the real performance of the model since it will be mostly determined by the values in the majority class. Metrics such as precision, recall, and the F1-score need be used to assess the performance in the different classes, individually.

Precision can be defined as the fraction of sources predicted as being from a certain class that are actually from that class:

\begin{equation}
Precision = \frac{TP}{TP+FP}
\end{equation}

The recall (also known as sensitivity or True Positive Rate; TPR) is the fraction of sources from a certain class that are predicted correctly:

\begin{equation}
Recall \equiv TPR = \frac{TP}{TP+FN}
\end{equation}

Both precision and recall have on the numerator the number of \textit{TP}. While in precision the denominator is the number of all predicted positive values, in recall it is the number of all real positive values. 
This means that precision reflects how reliable is the model when predicting if an element belongs to a particular class, while recall indicates how effectively the model recognises the elements from that class.
A combination of precision and recall can be given by the F1-score: 

\begin{equation}
F1 = \frac{2*Precision*Recall}{Precision+Recall}
\end{equation}

A lower value for either precision or recall will be reflected in this value. Therefore, this score is useful for identifying significant discrepancies between these two metrics.

To compute the Receiver Operating Characteristic (ROC) curve (see Figure~\ref{fig:roc_full_ds_dl}) we also use the False Positive Rate (FPR), which corresponds to the fraction of sources from the negative class that are incorrectly classified: 

\begin{equation}
FPR = \frac{FP}{FP+TN}
\end{equation}
\section{Rotation invariance of the final model}
\label{sec:rotation}

We explore the ability of the model to handle rotational and reflection symmetry, as source classification should not depend on any particular source orientation. \cite{khotanzad1990invariant} demonstrated that the Zernike moments exhibit inherent rotation invariance when extracted from a shape at different angles. In astronomy, this problem has been mostly addressed using CNNs to classify optical galaxies (e.g. \citet{dieleman2015rotation, khramtsov2022machine}, but recently received more interest in radio astronomy since orientation biases can be particularly problematic for the automatic classification of radio morphology sources into FRI and FRII in large surveys. \cite{ScaifeCNNequivariant2021} specifically designed CNNs to be group-equivariant and \cite{bowles2021} combined this with attention networks.

In order to test for this effect, we investigate the classification of the same source when seen from a variety of orientations and flips. We used only the  \texttt{PyBDSF} sources that belong to multi-component sources from the training set to inspect for this aspect; each image was randomly rotated and flipped as explained in the augmentation process. We did this four times, obtaining a total of 6,277 \texttt{PyBDSF} sources. The predictions for these sources were then calculated and compared. We calculated the standard deviation for the predictions of each group of 4 sources; the two sources with the most extreme variation had standard deviations between 0.05 and 0.1, but for the vast majority of the sources, the standard deviation was significantly below 0.01. We inspected the sources for which the probability showed higher differences, and the most extreme case corresponded to an example where part of the source was rotated outside of the image, with probabilities of being a MC ranging from 0.74 to 0.87. For the remaining sources, the differences seem less evident to the naked eye, with some emission obscured but still relevant for the classification. Nevertheless, for the majority of these sources, the predictions are skewed to one of the extremes, and so they do not translate into problems. 
For a threshold of 0.5, only 14 sources ended up with a mix of classifications, but those were very close to either above or below the 0.5 value. 
We can conclude that the algorithm is rotation-invariant, except if an important part of the source falls outside the cropped image for some rotation angles.


\bsp	
\label{lastpage}
\end{document}